\documentclass[aps,pra,twocolumn,superscriptaddress,notitlepage]{revtex4-1}
\usepackage{graphicx,amsfonts,amssymb,amsmath,hyperref,hypcap,enumerate,enumitem}
\usepackage{xcolor}
\usepackage{soul}
\usepackage{scalerel}
\usepackage{adjustbox,stmaryrd}
\usepackage{cancel}
\usepackage{csquotes}
\usepackage[utf8]{inputenc}
\usepackage{geometry}
\hypersetup{  colorlinks=true, linkcolor=blue, citecolor=blue, urlcolor=blue  }

\newcommand{\KITP}{Kavli Institute for Theoretical Physics, University of California, Santa Barbara, CA, 93106-4030}

\begin{document}

\title{Pseudospin density wave instability in two-dimensional electron bilayers }
\author{Jihang Zhu}
\affiliation{Condensed Matter Theory Center and Joint Quantum Institute, Department of Physics, University of Maryland,
College Park, Maryland 20742, USA}
\author{Tessa Cookmeyer}
\affiliation{\KITP}
\author{Sankar Das Sarma}
\affiliation{Condensed Matter Theory Center and Joint Quantum Institute, Department of Physics, University of Maryland,
College Park, Maryland 20742, USA}

\begin{abstract}
We investigate the instability of layer pseudospin paramagnetic (PSP) state to the formation of pseudospin density wave (PSDW) in two-dimensional (2D) electron bilayers, analogous to the formation of Overhauser spin density wave (SDW) in a single-layer 2D electron gas (2DEG) with spin 1/2. 
Our comprehensive study on phase diagrams, based on the self-consistent Hartree-Fock (HF) theory, reveals that the PSDW has a lower energy than both PSP and pseudospin ferromagnetic (PSF) states near the PSP-PSF phase transition boundary. 
When the two layers are populated by the same number of electrons, the PSDW momentum $Q_c \sim 2k_F$ near the PSP-PSDW boundary, where $k_F= (2\pi n)^{1/2}$ is the Fermi momentum characterized by the density in one of the two layers, and $Q_c$ decreases as the system transitions to the PSF regime.
Extending the HF study to the case of unequal layer densities, the PSP phase is unstable to PSDW for small density imbalances, with momentum $Q_c \sim k_{F,t} + k_{F,b}$, where $k_{F,t}$ and $k_{F,b}$ are Fermi momenta of top and bottom layers, respectively.
In PSDW regime, the ground state stability, defined by the energy difference between PSDW and the second lowest-energy state, is one order of magnitude lower than that in PSF regime, and decreases with increasing layer separation $d$.
Furthermore, incorporating RPA static screening with the Hubbard-type local field correction leads to disappearance of both SDW and PSDW phases, and pushes the phase boundaries of paramagnetic to ferromagnetic transitions to larger $r_s$ values.
Our study on PSDW in 2D electron bilayers is equally applicable to 2D hole bilayers.
The idea of pursuing PSDW is, in general, relevant across various 2D bilayer systems, not limited to the parabolic model that we investigate in this paper, and provides a new possibility of exploring novel coherent phases.
\end{abstract}



{\let\newpage\relax\maketitle}

\setcounter{page}{1} 

\section{Introduction}
\label{sec_intro}
A two-dimensional (2D) electron bilayer, abbreviated as an e-e bilayer in this paper as well, consists of two parallel 2D electron gas (2DEG) layers (in the $xy$-plane) separated by an out-of-plane distance $d$. Such an e-e bilayer, compared to a single-layer 2DEG with spin 1/2, gains complexity from an additional layer pseudospin, leading to interlayer coherence driven by the exchange interaction.
This interlayer coherence is associated with a spontaneous U(1) symmetry breaking in layer pseudospin.
Previous studies \cite{LZheng_doubleQW_1997, DasSarma_doubleQW_1998, JZ_interlayerCoherence_2024, Tessa_2Dbilayer_2024, XYbilayer_Girvin_2000} have investigated homogeneous interlayer coherence in e-e bilayers through the framework of Hartree-Fock (HF) mean-field theory, unraveling both ground-state behaviors and temperature-dependent phase transitions.
Other research \cite{SDS_2DbilayerSpin_1996, AHM_2DbilayerSpin_1998, Radtke_2DbilayerSpin_1996} on 2D bilayers have focused on various symmetry-broken phases in the spin sector, which spontaneously break SU(2) symmetry. In contrast, the pseudospin density wave (PSDW) discussed in this paper pertains to the layer pseudospin sector, with spontaneous U(1) symmetry breaking.

\begin{figure*}[!t]
\centering
\includegraphics[width=1.0\textwidth]{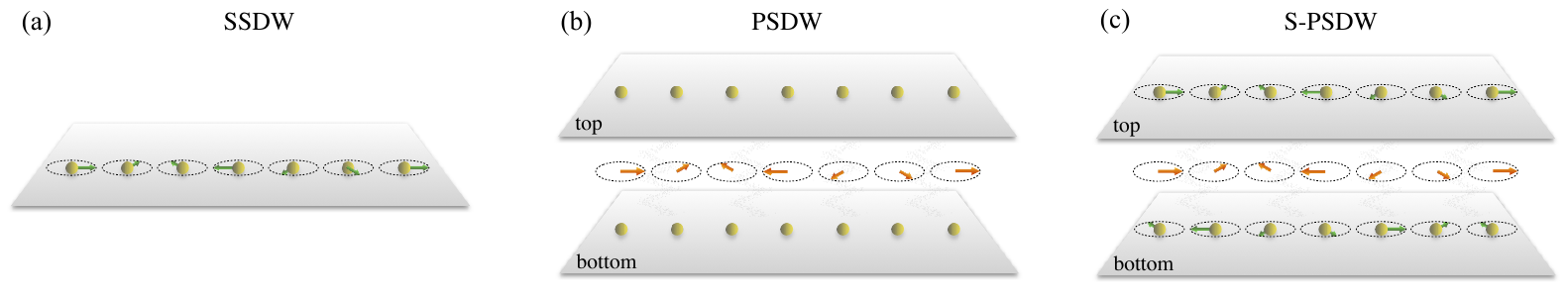}
\caption{\label{fig_schematic}
Schematic figures of spiral spin density wave (SSDW), pseudospin density wave (PSDW) and spin and pseudospin density wave (S-PSDW). 
(a) SSDW in a single-layer 2DEG, spin-up and spin-down electrons hybridize with a periodic phase factor.
(b) PSDW in a spinless e-e bilayer of equal layer densities. The layer pseudospin (orange arrows) points in the $xy$-plane and rotates periodically about $\hat{z}$ axis.
(c) S-PSDW in a spinful e-e bilayer, which specifically shows the case that intralayer SDW and interlayer PSDW have the same periodicity, which is not necessarily the case.
S-PSDW has an independent spin SU(2) rotational symmetry in each layer.}
\end{figure*}

In a single-layer 2DEG with spin 1/2, the original Overhauser instability theorem~\cite{Overhauser_SDW_1960, Overhauser_SDW_1962} states that the HF homogeneous paramagnetic state is unstable to the formation of either charge density wave (CDW) or spin density wave (SDW) for all electron densities. In the spiral spin density wave (SSDW) phase, a specific case of SDW, the HF energy is lowered by the hybridization of spin-up and spin-down species near the Fermi surface, with a phase factor periodic in real space. 
Figure~\ref{fig_schematic}(a) schematically illustrates a SSDW, in which the spin rotates periodically about the spin quantization axis $\hat{z}$.
Analogously, in an e-e bilayer, the layer pseudospin paramagnetic (PSP) state is unstable to the PSDW formation. The main difference between the layer PSDW and the Overhauser SDW is the dependency of the Coulomb potential in PSDW on the layer separation $d$, characterized by the factor $e^{-qd}$.

It is important to note that Overhauser's original argument for a long-range SDW for arbitrarily weak interactions can only hold at $T=0$ in 1D systems \cite{Kohn_SDW_1960}.
In 2D and 3D systems, the Overhauser SDW relies on the long-range nature of Coulomb interactions and can be negated by Thomas-Fermi static screening \cite{Fedders_correlation_1966}.
Despite these concerns, Overhauser SDWs in 2D and 3D electron gases are not completely ruled out, as rigorous dynamical screening, which is typically weaker than Thomas-Fermi static screening, might still rescue them. 
Additionally, CDWs and SDWs are more likely to occur in metals or semimetals with complex Fermi surfaces that satisfy Fermi surface nesting, or in materials with complex band structures that exhibit weaker dynamical screening, a topic we will discuss for future work in Sec.~\ref{sec_discussion}.

In this paper, we expand on previous homogeneous HF theory \cite{LZheng_doubleQW_1997, DasSarma_doubleQW_1998, JZ_interlayerCoherence_2024, Tessa_2Dbilayer_2024, XYbilayer_Girvin_2000} to provide a comprehensive study on the PSDW instability in 2D electron bilayers.
Our main focus is the case of equal layer densities where there are two main competing phases: the PSP, where the pseudospin is unpolarized and incoherent, and the pseudospin ferromagnet (PSF) where the pseudospin is polarized and oriented within the $xy$-plane. In Ref.~\cite{JZ_interlayerCoherence_2024}, the PSP (PSF) is referred to as $S_1$ ($S_2$) respectively.
In the phase diagram calculated by the self-consistent HF theory, we find a regime where the PSDW state has a lower energy than both PSP and PSF states.
The PSDW is stable near the PSP-PSF phase boundary calculated by previous homogeneous HF theory \cite{JZ_interlayerCoherence_2024}, across all layer separations $d$ and electron densities characterized by $r_s$, occupying a larger region within the PSP domain than in the PSF domain. 
In addition, as $d$ increases, the PSDW regime for a given $d$ spans a broader $r_s$ range.
The stability of PSDW state, characterized by its energy difference to the second lowest-energy state in the considered ground-state ansatz, is notably less robust (an order of magnitude smaller) than that in the PSF regime at large $r_s$ values, and the PSDW stability decreases with increasing $d$. 
The PSDW momentum $Q_c$ is approximately $2k_F$ near the PSDW-PSP boundary, decreases towards the PSDW-PSF boundary and drops to zero quickly when approaching the PSF phase. This property is analogous to the Overhauser SDW momentum $Q_c^{\rm SDW}$ as a function of $r_s$, which begins around $2k_F$ at low $r_s$ values and decreases with increasing $r_s$.
Furthermore, the RPA static screening, self-consistently included in our HF calculations, eliminates all coherent phases, including the spin ferromagnetic and SDW phases (the $d=0$ limit of our bilayer model), as well as PSF and PSDW phases for all $d$ values, leaving only the spin and pseudospin paramagnetic phase in the phase diagram. 
After adding Hubbard-type local field corrections, however, the spin ferromagnetic and PSF phases re-appear, though their phase boundaries shift to higher $r_s$ values.

Extending to the case of unequal layer densities, the fate of PSDW is similar to that of equal layer densities: PSDW phase occupies the regime near the phase boundary of pseudospin incoherent ($S_1'$ in Ref.~\cite{JZ_interlayerCoherence_2024}) and pseudospin coherent ($S_\xi$ in Ref.~\cite{JZ_interlayerCoherence_2024}) phases, with momentum $Q_c \sim k_{F,t}+k_{F,b}$ near the pseudospin-incoherent-PSDW boundary and decreases towards the pseudospin-coherent-PSDW boundary.
For a fixed $d$, the PSDW only occurs for a small density imbalance since the intralayer exchange interaction tends to polarize all electrons to one of the layers if either layer imbalance $m$ or average inter-electron distance $\bar{r}_s$ is large.

The detection of PSDW proposed in this paper is likely achievable through magneto-transport experiments \cite{Ando_QOscill_1974}, which are adept at assessing Fermi surface properties, such as the number of different orbits and their contours \cite{Smrcka_CyclotronMass_1995, Kaganov_Lifshits_1979}.
Compared to the incoherent state (PSP), interlayer coherent states (PSF and PSDW) are distinguishable from Shubnikov–de Haas (SdH) oscillations, through the variation of the number of Fermi surfaces with electron density.
The minigap (gap) induced by interlayer coherence in PSDW (PSF) manifests as anomalies in zero-field conductivity \cite{Cole_tiltedSi_1977, Tsui_tiltedSi_1978} measured as a function of electron concentration, owing to van Hove singularities at band edges. 
Optical absorption techniques can quantify the minigap or gap size \cite{Sesselmann_OpticalGap_1979, Kamgar_OpticalGap_1980}.
Further differentiation between PSF and PSDW requires detailed SdH oscillation analyses relative to electron doping levels, which provides insights into the Fermi surface geometry.
Unlike the isotropic Fermi surface of PSF, PSDW features a non-isotropic Fermi surface due to broken translational and rotational symmetries.
These analyses remain valid despite non-negligible interlayer tunneling \cite{Matheson_tiltedSi_1982}.
Anti-crossing band structures similar to PSDW have been studied in double quantum wells subject to in-plane magnetic fields \cite{Simmons_dQWinBpara_1994, Simmons_dQWinBpara_1995, Kurobe_dQWinBpara_1994, Lyo_dQWinBpara_1994, Lyo_dQWinBpara_1995} and in electron inversion layers on tilted Si surfaces \cite{Ando_minigap_review_1982, Cole_tiltedSi_1977, Sham_tiltedSi_1978, Tsui_tiltedSi_1978, Matheson_tiltedSi_1982}, as well as in the 2DEG with broken translational symmetry \cite{Winkler_1Dsuperlattice_1989}.
Therefore, the experimental detection of PSDW is anticipated to be similar to these well-studied systems.
Additionally, the band structures featuring minigaps and detailed Fermi surfaces can be directly imaged using angle-resolved photoemission spectroscopy (ARPES).

This paper is organized as follows.
Section~\ref{sec_2band} introduces the HF Hamiltonian and self-consistent equations that breaks both translational and rotational symmetries, assuming equal electron populations in the two layers.
Section~\ref{sec_SDW} examines the Overhauser SDW by setting the interlayer distance, $d$, to zero in the previously discussed model. 
The HF phase diagram, parameterized by $r_s$ and $d$, is presented in Sec.~\ref{sec_HFphase_unscreened}.
To assist in future experiments, Sec.~\ref{sec_unequal_density} extends the analysis of PSDW and provides the HF phase diagram to scenarios of unequal layer densities.
Section~\ref{sec_RPA} investigates the impact of screening on spin and pseudospin coherences, in which both the primitive RPA static screening and the ones corrected by Hubbard-type local field factors are self-consistently taken into account in the HF calculations.
Lastly in Sec.~\ref{sec_discussion}, we comment on the impact of a finite interlayer tunneling on PSDW, as well as the relevance of PSDW to other complex 2D systems, for example graphene- and TMD-based 2D materials with valley degree of freedom.
Additional discussions on the spinful bilayer model are included in Appendix~\ref{supp_4band}.

Although our study utilizes the 2D bilayer of parabolic electron gases as an example to conduct calculations, the findings on PSDW instability is general, extending beyond this specific electronic structure, and equally applicable to 2D hole systems.

\section{Pseudospin density wave instability}
\label{sec_para_instability}
In this section, we discuss the PSDW instability in an e-e bilayer of equal layer densities. We focus on the spinless PSDW model, which is justified by the decoupling of spin and pseudospin sectors in our model.
This decoupling allows us to discuss pseudospin phases independently of spin, especially in the low density regime that spin is polarized.
The spinless PSDW model is schematically shown in Fig.~\ref{fig_schematic}(b).
The SU(2) symmetric Overhauser SDW theory in a single-layer 2DEG, depicted in Fig.~\ref{fig_schematic}(a), is recovered in the $d = 0$ limit of this model.

\subsection{The model}
\label{sec_2band}
Consider a Hamiltonian $\mathcal{H}$ with two-fold degrees of freedom denoted by $s$ and $\bar{s}$, the translational and rotational symmetries are broken by the single momentum $\mathbf{Q}$,
\begin{align}
\label{Eq_Hamil}
\mathcal{H} = & \sum\limits_{\mathbf{k}} \frac{\hbar^2 |\mathbf{k}+\mathbf{Q}/2|^2}{2m^*} c^\dagger_{s, \mathbf{k}+\frac{\mathbf{Q}}{2}} c_{s, \mathbf{k}+\frac{\mathbf{Q}}{2}} \\
+ & \sum\limits_{\mathbf{k}} \frac{\hbar^2 |\mathbf{k}-\mathbf{Q}/2|^2}{2m^*} c^\dagger_{\bar{s}, \mathbf{k}-\frac{\mathbf{Q}}{2}} c_{\bar{s}, \mathbf{k}-\frac{\mathbf{Q}}{2}} \nonumber\\
+ & \frac{1}{2A} 
\sum\limits_{\mathbf{k},\mathbf{k}',\mathbf{q}} V^0_\mathbf{q} c^\dagger_{s, \mathbf{k}+\frac{\mathbf{Q}}{2}+\mathbf{q}} c^\dagger_{s, \mathbf{k}'+\frac{\mathbf{Q}}{2}-\mathbf{q}} c_{s,\mathbf{k}'+\frac{\mathbf{Q}}{2}} c_{s,\mathbf{k}+\frac{\mathbf{Q}}{2}} \nonumber\\
+ & \frac{1}{2A} 
\sum\limits_{\mathbf{k},\mathbf{k}',\mathbf{q}} V^0_\mathbf{q} c^\dagger_{\bar{s}, \mathbf{k}-\frac{\mathbf{Q}}{2}+\mathbf{q}} c^\dagger_{\bar{s}, \mathbf{k}'-\frac{\mathbf{Q}}{2}-\mathbf{q}} c_{\bar{s},\mathbf{k}'-\frac{\mathbf{Q}}{2}} c_{\bar{s},\mathbf{k}-\frac{\mathbf{Q}}{2}} \nonumber\\
+ & \frac{1}{A} \sum\limits_{\mathbf{k},\mathbf{k}',\mathbf{q}} V^d_\mathbf{q} c^\dagger_{s, \mathbf{k}+\frac{\mathbf{Q}}{2}+\mathbf{q}} c^\dagger_{\bar{s}, \mathbf{k}'-\frac{\mathbf{Q}}{2}-\mathbf{q}} c_{\bar{s},\mathbf{k}'-\frac{\mathbf{Q}}{2}} c_{s,\mathbf{k}+\frac{\mathbf{Q}}{2}}. \nonumber
\end{align}
Without loss of generality, $\mathbf{Q} = Q\hat{x}$ is chosen to be in the $x$-axis.
For layer pseudospins, $s = t$ and $\bar{s} = b$.
$m^*$ is the effective mass, $A$ is the sample area and $\epsilon_b$ is the dielectric constant of the surrounding dielectric environment. 
$V^d_\mathbf{q} = 2\pi e^2 e^{-qd}/\epsilon_b q$ is the interlayer-separation-dependent Coulomb potential: $d=0$ for intralayer interactions and $d \neq 0$ for interlayer interactions.
The HF Hamiltonian $H(\mathbf{k})$ with basis spinor $(c_{s,\mathbf{k}+\mathbf{Q}/2}, c_{\bar{s},\mathbf{k}-\mathbf{Q}/2})^T$ is
\begin{equation}
\label{eq_hamil_2band}
\begin{split}
H(\mathbf{k}) = 
\begin{pmatrix}
\varepsilon_{s, \mathbf{k}+\mathbf{Q}/2} & -\Delta_\mathbf{k} \\
-\Delta^*_\mathbf{k} & \varepsilon_{\bar{s}, \mathbf{k}-\mathbf{Q}/2}
\end{pmatrix},
\end{split}
\end{equation}
and its quasiparticle eigenenergies and eigenvectors are
\begin{equation}
\label{Eq_eigens}
\begin{gathered}
\varepsilon_{\pm,\mathbf{k}} = \frac{1}{2}( \varepsilon_{s, \mathbf{k}+\frac{\mathbf{Q}}{2}} + \varepsilon_{\bar{s}, \mathbf{k}-\frac{\mathbf{Q}}{2}} )
\pm \sqrt{ \xi^2_\mathbf{k} + \Delta_\mathbf{k}^2 }, \\
\begin{pmatrix}
+,\mathbf{k} \\
-,\mathbf{k}
\end{pmatrix}
=
\begin{pmatrix}
u_\mathbf{k} & -v_\mathbf{k} \\
v^*_\mathbf{k} & u^*_\mathbf{k}
\end{pmatrix}
\begin{pmatrix}
c_{s, \mathbf{k}+\frac{\mathbf{Q}}{2}} \\
c_{\bar{s}, \mathbf{k}-\frac{\mathbf{Q}}{2}}
\end{pmatrix},
\end{gathered}
\end{equation}
where
\begin{equation}
\label{Eq_eps}
\begin{gathered}
\varepsilon_{s, \mathbf{k}+\frac{\mathbf{Q}}{2}} = \frac{\hbar^2|\mathbf{k}+\mathbf{Q}/2|^2}{2m^*} - \frac{1}{A}\sum\limits_{\mathbf{k}'} V^0_{\mathbf{k}'-\mathbf{k}} \rho_{ss}(\mathbf{k}'), \\
\varepsilon_{\bar{s}, \mathbf{k}-\frac{\mathbf{Q}}{2}} = \frac{\hbar^2|\mathbf{k}-\mathbf{Q}/2|^2}{2m^*} - \frac{1}{A}\sum\limits_{\mathbf{k}'} V^0_{\mathbf{k}'-\mathbf{k}} \rho_{\bar{s}\bar{s}}(\mathbf{k}'), \\
\xi_\mathbf{k} = \frac{1}{2}( \varepsilon_{s, \mathbf{k}+\frac{\mathbf{Q}}{2}} - \varepsilon_{\bar{s}, \mathbf{k}-\frac{\mathbf{Q}}{2}}), \\
\Delta_{\mathbf{k}} = \frac{1}{A} \sum\limits_{\mathbf{k}'} V^d_{\mathbf{k}'-\mathbf{k}} \rho_{s\bar{s}}(\mathbf{k}').
\end{gathered}
\end{equation}
$\Delta_{\mathbf{k}}$ is the PSDW order parameter with momentum $\mathbf{Q}$. Order parameters with larger momenta $2\mathbf{Q}$, $3\mathbf{Q}$, and so on, are neglected.
$\rho_{ss}(\mathbf{k})$, $\rho_{\bar{s}\bar{s}}(\mathbf{k})$ and $\rho_{s\bar{s}}(\mathbf{k})$ are density matrix elements defined as
\begin{equation}
\begin{split}
\rho_{ss}(\mathbf{k}) = \langle c^\dagger_{s, \mathbf{k}+\frac{\mathbf{Q}}{2}} c_{s, \mathbf{k}+\frac{\mathbf{Q}}{2}} \rangle,  \\
\rho_{\bar{s}\bar{s}}(\mathbf{k}) = \langle c^\dagger_{\bar{s}, \mathbf{k}-\frac{\mathbf{Q}}{2}} c_{\bar{s}, \mathbf{k}-\frac{\mathbf{Q}}{2}} \rangle, \\
\rho_{s\bar{s}}(\mathbf{k}) = \langle c^\dagger_{\bar{s}, \mathbf{k}-\frac{\mathbf{Q}}{2}} c_{s, \mathbf{k}+\frac{\mathbf{Q}}{2}} \rangle.
\end{split}
\end{equation}
The expectations are explicitly
\begin{equation}
\label{Eq_expectations}
\begin{split}
\langle c^\dagger_{s, \mathbf{k}+\frac{\mathbf{Q}}{2}} c_{s, \mathbf{k}+\frac{\mathbf{Q}}{2}} \rangle
&= |v_{\mathbf{k}}|^2
f_{-,\mathbf{k}} + |u_{\mathbf{k}}|^2 f_{+,\mathbf{k}},  \\
\langle c^\dagger_{\bar{s}, \mathbf{k}-\frac{\mathbf{Q}}{2}} c_{\bar{s}, \mathbf{k}-\frac{\mathbf{Q}}{2}} \rangle 
&= |u_{\mathbf{k}}|^2 f_{-,\mathbf{k}} + |v_{\mathbf{k}}|^2 f_{+,\mathbf{k}},  \\
\langle c^\dagger_{\bar{s}, \mathbf{k}-\frac{\mathbf{Q}}{2}} c_{s, \mathbf{k}+\frac{\mathbf{Q}}{2}} \rangle 
&= u_{\mathbf{k}} v^*_{\mathbf{k}} (f_{-,\mathbf{k}} - f_{+,\mathbf{k}}),  \\
\end{split}
\end{equation}
where $f_{\pm,\mathbf{k}} \equiv f(\varepsilon_{\pm,\mathbf{k}}-\mu)$ is the Fermi-Dirac distribution function.
The self-consistent equations to be solved are
\begin{align}
\label{Eq_scEqs_PSDW}
\xi_{\mathbf{k}} &= \frac{\hbar^2}{4m^*} \Big( |\mathbf{k}+\frac{\mathbf{Q}}{2} |^2 - |\mathbf{k}-\frac{\mathbf{Q}}{2} |^2 \Big) \nonumber\\
& \quad + \frac{1}{A} \sum\limits_{\mathbf{k}'} V^0_{\mathbf{k}'-\mathbf{k}} \frac{\xi_{\mathbf{k}'}}{\sqrt{\xi_{\mathbf{k}'}^2 + \Delta_{\mathbf{k}'}^2}} (f_{-,\mathbf{k}'} - f_{+,\mathbf{k}'}), \nonumber\\
\Delta_{\mathbf{k}} &= \frac{1}{2A} \sum\limits_{\mathbf{k}'} V^d_{\mathbf{k}'-\mathbf{k}} \frac{\Delta_{\mathbf{k}'}}{\sqrt{\xi_{\mathbf{k}'}^2 + \Delta_{\mathbf{k}'}^2}} (f_{-,\mathbf{k}'} - f_{+,\mathbf{k}'}).
\end{align}
The HF energy per electron, $\varepsilon_{\rm tot}$, is the sum of the kinetic energy $\varepsilon_{\rm kin}$ and the exchange energy $\varepsilon_{\rm x}$, which includes intralayer $\varepsilon_{\rm x}^{\rm intra}$ and interlayer $\varepsilon_{\rm x}^{\rm inter}$ contributions:
\begin{equation}
\begin{split}
\varepsilon_{\rm tot} &= \varepsilon_{\rm kin} + \varepsilon_{\rm x}^{\rm intra} + \varepsilon_{\rm x}^{\rm inter}, \\
\varepsilon_{\rm kin} &= \frac{\hbar^2}{2m^*N} \sum\limits_{\mathbf{k}} \Big( \big|\mathbf{k} + \frac{\mathbf{Q}}{2}\big|^2 \rho_{ss}(\mathbf{k})
\\
& \qquad \qquad \quad \ + \big|\mathbf{k} - \frac{\mathbf{Q}}{2} \big|^2 \rho_{\bar{s} \bar{s}}(\mathbf{k}) \Big), \\
\varepsilon_{\rm x}^{\rm intra} &= -\frac{1}{2AN} \sum\limits_{\mathbf{k}, \mathbf{k}'} V^0_{\mathbf{k}'-\mathbf{k}} \big[ \rho_{ss}(\mathbf{k}') \rho_{ss}(\mathbf{k}) \\
& \qquad \qquad \qquad \qquad + \rho_{\bar{s} \bar{s}}(\mathbf{k}') \rho_{\bar{s} \bar{s}}(\mathbf{k}) \big],
\\
\varepsilon_{\rm x}^{\rm inter} 
&= -\frac{1}{AN} \sum\limits_{\mathbf{k}, \mathbf{k}'} V^d_{\mathbf{k}'-\mathbf{k}}
\rho_{s\bar{s}}(\mathbf{k}') \rho^*_{s\bar{s}}(\mathbf{k}),
\end{split}
\end{equation}
where $N=nA$ is the total number of electrons in the system. 
The Hartree energy vanishes because of equal layer densities.

The convergent HF energy as a function of $\mathbf{Q}$ depends on the initial conditions of the self-consistent equations Eq.~(\ref{Eq_scEqs_PSDW}).
For most parameter sets ($r_s,d$) in the phase diagram, the minimum HF energy is found by initializing Eq.~(\ref{Eq_scEqs_PSDW}) using energies of the PSF state, i.e.,

\begin{align}
\label{Eq_HF_init}
\xi^{(0)}_{\mathbf{k}} &= \frac{\hbar^2}{4m^*} \Big( |\mathbf{k}+\frac{\mathbf{Q}}{2} |^2 - |\mathbf{k}-\frac{\mathbf{Q}}{2} |^2 \Big) \nonumber\\
& - \frac{e^2k_F}{2\pi \epsilon_b} \Big[ f_{\rm 2D}\big(\frac{|\mathbf{k}+\mathbf{Q}/2|}{k_F}\big) - f_{\rm 2D}\big(\frac{|\mathbf{k}-\mathbf{Q}/2|}{k_F}\big) \Big], \nonumber\\
\Delta^{(0)}_{\mathbf{k}} &= 
\frac{e^2k_F}{4\pi \epsilon_b} \Big[I\big( \frac{|\mathbf{k}+\mathbf{Q}/2|}{k_F}, k_Fd \big) \nonumber\\
& \qquad \quad + I\big( \frac{|\mathbf{k}-\mathbf{Q}/2|}{k_F}, k_Fd \big)\Big],
\end{align}
where $u_{\mathbf{k}} = v_{\mathbf{k}} = 1/\sqrt{2}$ and $f_{-,\mathbf{k}} = 1, f_{+,\mathbf{k}} = 0$ are used. Functions $I(x,k_Fd)$ and $f_{\rm 2D}(x)$ \cite{JZ_interlayerCoherence_2024} are defined as:
\begin{equation}
\begin{split}
I(x,k_Fd) &= \int_0^1 dy y \int_0^{2\pi} d\theta \frac{e^{-k_Fd \sqrt{x^2+y^2-2xy\cos \theta}}}{\sqrt{x^2 + y^2-2xy\cos \theta}}, \\
f_{\rm 2D}(x) &= \frac{1}{4}I(x, k_Fd=0) \\
&= \begin{cases} 
E(x), & x \leq 1, \\
x \left[ E\left(\frac{1}{x}\right) - \left(1 - \frac{1}{x^2}\right) K\left(\frac{1}{x}\right) \right], & x \geq 1.
\end{cases}
\end{split}
\end{equation}
$K(x)$ and $E(x)$ are the complete elliptic integral of the first and the second kind, respectively. 
Near the PSDW-PSP phase transition boundary, the minimum HF energy is obtained by starting with a small perturbation to the PSP state, although this energy only slightly differs from that obtained by starting from the PSF state using Eq.~(\ref{Eq_HF_init}). 
In the following presented results, we only consider the PSDW state to have lower energy than PSP or PSF state if the energy difference is greater than $\sim 0.001$ Ry$^*$. The scale of $k$-grid in these calculations is chosen to be $\sim 0.036k_F$ and the momentum cutoff $\sim 3.6k_F$. The self-consistent equations are solved until convergence when the $k$-average of order parameter $\bar{\Delta}_k < 10^{-4}$ to $10^{-5}$ Ry$^*$.
Here Ry$^*$ is the effective Rydberg
\begin{equation}
{\rm Ry}^* = \frac{e^2}{2a^* \epsilon_b} = \frac{\hbar^2}{2m^* (a^*)^2}.
\end{equation}
$a^*=\epsilon_b \hbar^2/e^2m^*$ is the effective Bohr radius. In GaAs-AlGaAs double quantum well, $\epsilon_b = 12.5$, $m^*=0.07m_e$ \cite{shklovskii2013}, $a^* = 98.3 \mathring{A}$ and 
Ry$^* \approx 5.5$ meV.

\subsection{The $d=0$ limit --- the Overhauser SDW}
\label{sec_SDW}
\begin{figure}[!t]
\centering
\includegraphics[width=1.0\columnwidth]{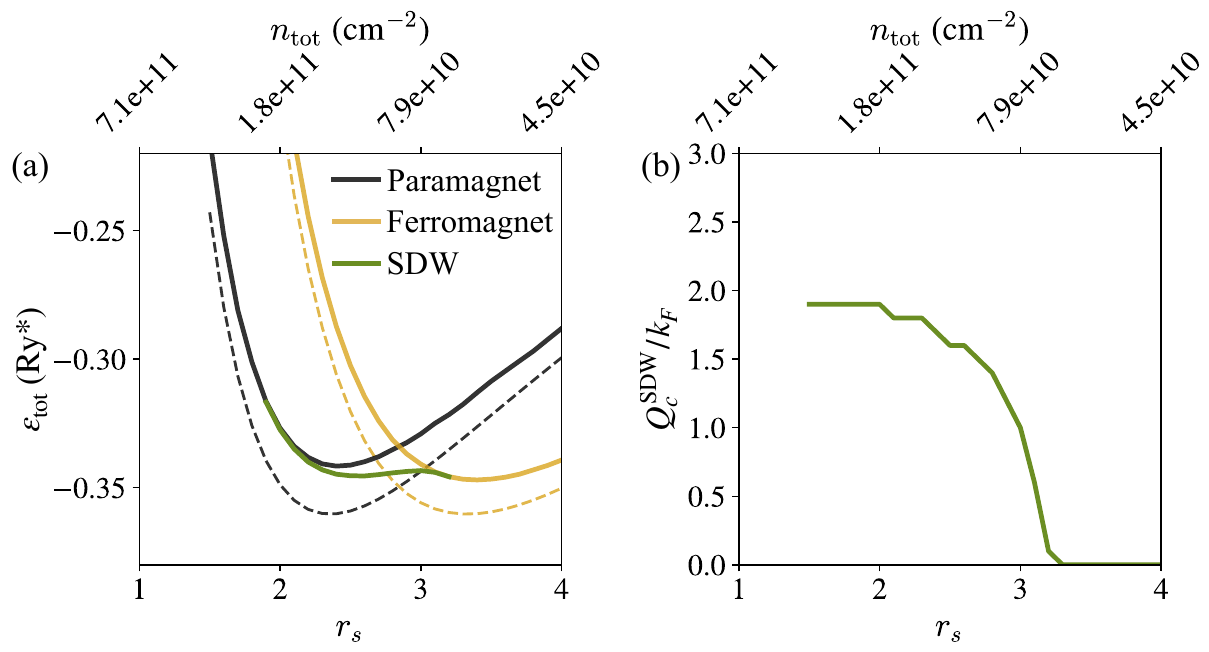}
\caption{\label{fig_SU2} {
SDW in a single-layer 2DEG.
(a) HF energies per electron as a function of $r_s$ for paramagnetic (black), ferromagnetic (yellow) and SDW (green) states. 
The dashed lines are calculated using the exact formula Eq.~(\ref{Eq_eptot_SU2_analy}) and the solid lines, slightly higher in energy, are calculated by numerically summing over $k$-space occupied states using Eq.~(\ref{Eq_eptot_k}). The SDW state exhibits lower energy at intermediate densities, $r_s \in [1.9, 3.2]$, with the most pronounced instability near the paramagnetic to ferromagnetic transition at $r_s \sim 2.8$, as indicated by the energy difference in (a).
(b) The SDW momentum, $Q^{\rm SDW}_c$, plotted as a function of $r_s$. $Q^{\rm SDW}_c$ is approximately $2k_F$ at high densities ($r_s \lesssim 2$), decreases as the density decreases ($r_s$ increases) and drops to zero rapidly as $r_s$ approaches the critical value $\sim 3.2$, where $k_F \equiv k_F^P = (2\pi n)^{1/2}$.
In these calculations, the scale of $k$-grid is chosen to be $\sim 0.036k_F$. The step-like features in (b) are a result of finite number of $Q$ gridding in our calculation.
  }}
\end{figure}

The $d=0$ limit of the model described in Sec.~\ref{sec_2band}, setting spins $s = \downarrow$ and $\bar{s} = \uparrow$, recovers the Overhauser SDW theory with SU(2) symmetry in a single-layer 2DEG.
The HF energies per electron of the SU(2) paramagnetic ($\varepsilon_{\rm tot}^{\rm P}$ ) and ferromagnetic ($\varepsilon_{\rm tot}^{\rm F}$) states are, respectively,
\begin{equation}
\label{Eq_eptot_SU2_analy}
\begin{split}
\frac{\varepsilon_{\rm tot}^{\rm P}}{\text{Ry}^*} &= \pi(a^*)^2 n - \frac{8\sqrt{2}a^*}{3\sqrt{\pi}} n^{1/2}, \\
\frac{\varepsilon_{\rm tot}^{\rm F}}{\text{Ry}^*} &= 2\pi(a^*)^2 n - \frac{16a^*}{3\sqrt{\pi}} n^{1/2},
\end{split}
\end{equation}
where $n$ is the total electron density. $\varepsilon_{\rm tot}^{\rm P}$ matches the energy of the PSP state ($S_1$ in Ref.~\cite{JZ_interlayerCoherence_2024}), and $\varepsilon_{\rm tot}^{\rm F}$ corresponds to the $d=0$ limit of the PSF state ($d=0$ limit of $S_2$ in Ref.~\cite{JZ_interlayerCoherence_2024}). 
Figure~\ref{fig_SU2}(a) shows $\varepsilon_{\rm tot}^{\rm P}$ (black dashed line) and $\varepsilon_{\rm tot}^{\rm F}$ (yellow dashed line) as a function of $r_s = \sqrt{2/\pi n}/a^*$. The SU(2) paramagnetic to ferromagnetic transition occurs at $r_s \sim 2.8$, agreeing with the critical density for the $S_1$ to $S_2$ phase transition in the $d=0$ limit (Fig.~2 in Ref.~\cite{JZ_interlayerCoherence_2024}).
Equivalently, $\varepsilon_{\rm tot}^{\rm P}$ and $\varepsilon_{\rm tot}^{\rm F}$ can be expressed as a summation over occupied states in $k$-space,
\begin{equation}
\label{Eq_eptot_k}
\begin{split}
\frac{\varepsilon_{\rm tot}^{\rm P}}{\text{Ry}^*} &= \frac{2(a^*)^2}{N} \sum\limits_{k \leq k^P_F} k^2 - \frac{4\pi a^*}{NA} \sum\limits_{k,k' \leq k^P_F} \frac{1}{|\mathbf{k} - \mathbf{k}'|}, \\
\frac{\varepsilon_{\rm tot}^{\rm F}}{\text{Ry}^*} &= \frac{(a^*)^2}{N} \sum\limits_{k \leq k^F_F} k^2 - \frac{2\pi a^*}{NA} \sum\limits_{k,k' \leq k^F_F} \frac{1}{|\mathbf{k} - \mathbf{k}'|},
\end{split}
\end{equation}
where $k^P_F = (2\pi n)^{1/2}$ and $k^F_F = (4\pi n)^{1/2}$ are Fermi momenta of paramagnetic and ferromagnetic states, respectively. Energies calculated using Eq.(\ref{Eq_eptot_k}) are plotted as solid lines in Fig.~\ref{fig_SU2}(a), which are slightly higher in energy than those calculated using the exact formula Eq.~(\ref{Eq_eptot_SU2_analy}) but accurately capture the critical $r_s$ of the paramagnetic to ferromagnetic transition.

\begin{figure}[!t]
\centering
\includegraphics[width=1.0\columnwidth]{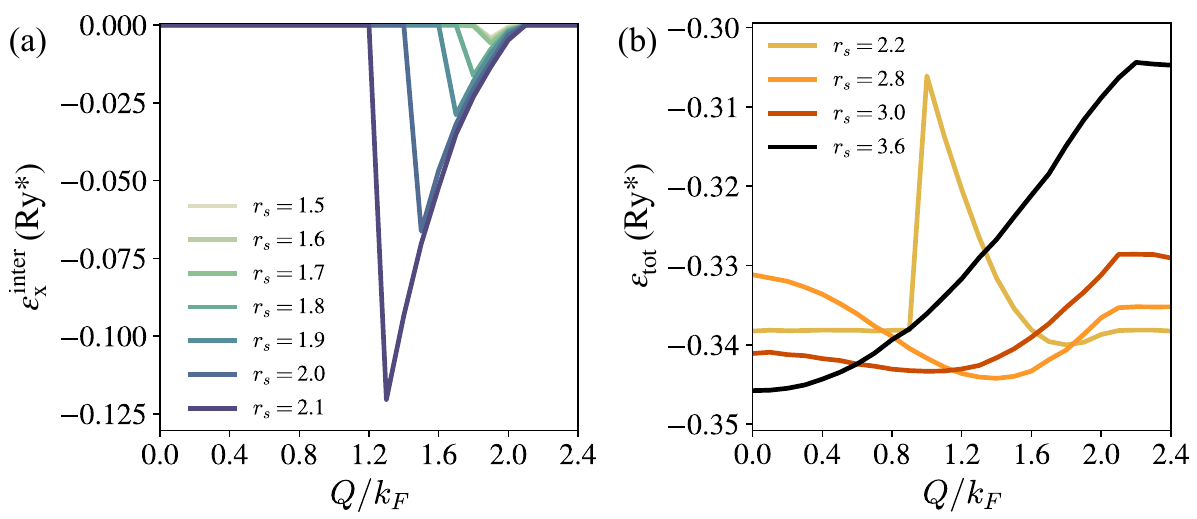}
\caption{\label{fig_epxinter_SU2} {
SDW in a single-layer 2DEG.
(a) The exchange energy $\varepsilon_x^{\rm inter}$, which quantifies the coherence between opposite spins, as a function of $Q$ for several small $r_s$ values. For $r_s \lesssim 1.8$, $\varepsilon_x^{\rm inter}$ is significantly smaller than the energy scale of $\varepsilon_{\rm tot}$ in Fig.~\ref{fig_SU2}(a), and therefore coherence can be easily inundated by correlations.
(b) The HF energy $\varepsilon_{\rm tot}$ versus $Q$ for $r_s=2.2, 2.8, 3.0$ and $3.6$. In the SDW regime, $r_s \in [1.9, 3.2]$, the energy difference between the SDW and the second lowest-energy state increases and then decreases with increasing $r_s$, a trend which can be seen in Fig.~\ref{fig_SU2}(a) as well.}
}
\end{figure}

\begin{figure*}[!htb]
\centering
\includegraphics[width=1.0\textwidth]{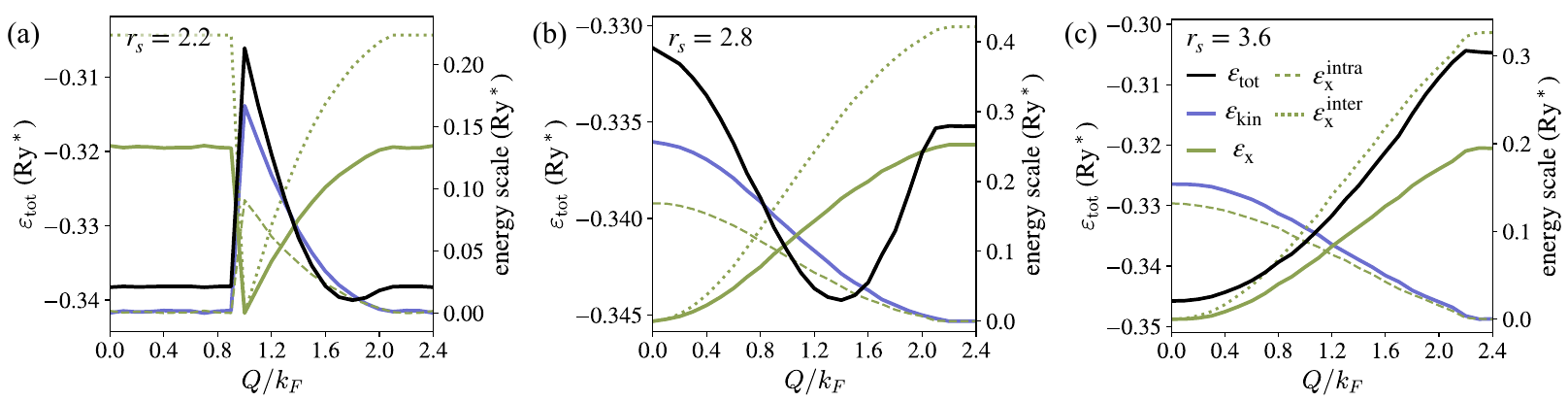}
\caption{\label{fig_eps_SU2} {
All energy components of the Overhauser SDW HF energy $\varepsilon_{\rm tot}$ (black solid lines): the kinetic energy $\varepsilon_{\rm kin}$ (blue solid lines) and the exchange energy $\varepsilon_{\rm x}$ (green solid lines), including the one within the same spin $\varepsilon_{\rm x}^{\rm intra}$ (green dashed lines) and the one captures the coherence between opposite spins $\varepsilon_{\rm x}^{\rm inter}$ (green dotted lines), for (a) $r_s = 2.2$, (b) $r_s = 2.8$ and (c) $r_s = 3.6$. The dominance of $\varepsilon_{\rm x}^{\rm inter}$ at high $r_s$ values leads to the stabilization of the ferromagnetic state ($Q^{\rm SDW}_c=0$), for example in (c).
In each figure, the $y$-axis on the left measures $\varepsilon_{\rm tot}$, and the $y$-axis on the right scales other energy components ($\varepsilon_{\rm kin}$, $\varepsilon_{\rm x}$, $\varepsilon_{\rm x}^{\rm intra}$ and $\varepsilon_{\rm x}^{\rm inter}$) which are all offset for clarity and comparative convenience in the same figure. 
  }}
\end{figure*}

The green line in Fig.~\ref{fig_SU2}(a) plots the energy of the SU(2) SDW, which is lower than both paramagnetic and ferromagnetic states for intermediate electron densities, $r_s \in [1.9, 3.2]$. The corresponding SDW momentum $Q^{\rm SDW}_c$, as a function of $r_s$, is shown in Fig.~\ref{fig_SU2}(b). At high densities ($r_s \lesssim 2$), $Q^{\rm SDW}_c$ is approximately $2k_F$, where $k_F \equiv k_F^P=(2\pi n)^{{1/2}}$. As the density decreases ($r_s$ increases), $Q^{\rm SDW}_c$ decreases and drops to zero rapidly as $r_s$ approaches the critical value $\sim 3.2$ \cite{Zhang_SDW_2008}. 
The step-like features in Fig.~\ref{fig_SU2}(b) are a result of finite number of $Q$ gridding in our calculations.

In the high-density limit, the correlation effects are relatively weak and the exchange-driven instability, the Overhauser SDW here, should be stable. Our calculations shown in Fig.~\ref{fig_SU2}(a), however, indicate that for $r_s \lesssim 1.9$, there is no obvious energy benefit from the SDW formation. This is because the exchange energy gain from the single Slater determinant SDW state becomes exponentially small \cite{Giuliani_SDW_2008, Kurth_SDW_SDFT_2009, Delyon_SDW_2015, Gontier_SDW_2019, Christiansen2023} at low $r_s$ values and can be easily inundated by correlations. 
The exchange energy $\varepsilon_x^{\rm inter}$, which quantifies the coherence between opposite spins, is plotted as a function of $Q$ for several small $r_s$ values in Fig.~\ref{fig_epxinter_SU2}(a).
For $r_s \lesssim 1.8$, $\varepsilon_{\rm x}^{\rm inter}$ is significantly smaller than the energy scale of $\varepsilon_{\rm tot}$ in Fig.~\ref{fig_SU2}(a).
Figure~\ref{fig_epxinter_SU2}(b) shows $\varepsilon_{\rm tot}$ versus $Q$ for $r_s=2.2, 2.8, 3.0$ and $3.6$. In the SDW regime, $r_s \in [1.9, 3.2]$, the energy difference between the SDW and the second lowest-energy state increases and then decreases with increasing $r_s$, a trend which can be seen in Fig.~\ref{fig_SU2}(a) as well. 
The SDW instability is most pronounced near the paramagnetic to ferromagnetic transition at $r_s \sim 2.8$ (Fig.~\ref{fig_SU2}(a)) when $\varepsilon_{\rm tot}(Q=0)$ approximates $\varepsilon_{\rm tot}(Q=2k_F)$ in Fig.~\ref{fig_epxinter_SU2}(b). This property agrees with findings from previous studies in 3DEG \cite{FGEich_SDW_3DEG, Baguet_3DEG_2014, Zhang_SDW_2008}.

In Fig.~\ref{fig_eps_SU2}, we show all energy components---the kinetic energy $\varepsilon_{\rm kin}$, the exchange energy $\varepsilon_{\rm x}$ including both intra-spin ($\varepsilon_{\rm x}^{\rm intra}$) and inter-spin ($\varepsilon_{\rm x}^{\rm inter}$) components---for $r_s=2.2, 2.8$ and $3.6$. 
At low $r_s$ (high densities), inter-spin coherence is confined to a small region in momentum space near $Q \sim 2k_F$. At high $r_s$, $\varepsilon_{\rm x}^{\rm inter}$ dominates and reaches its minimum at zero momentum, thus favoring the ferromagnetic state ($Q^{\rm SDW}_c=0$), as for example depicted in Fig.~\ref{fig_eps_SU2}(c) for $r_s=3.6$.
It should be noted that the sailboat-like energy profiles at $r_s = 2.2$ in Figs.~\ref{fig_epxinter_SU2}(b) and \ref{fig_eps_SU2}(a) directly result from the specific HF initialization condition, which is chosen to be Eq.~(\ref{Eq_HF_init}) in these figures. Such an initial condition results in the peak in the kinetic energy $\varepsilon_{\rm tot}$ for $Q < 2k_F$  near the PSP-PSDW phase transition.

\subsection{The HF phase diagram}
\label{sec_HFphase_unscreened}
The spinless model of PSDW in 2D electron bilayers closely resembles the SDW in a single-layer 2DEG, except that the Coulomb interaction in PSDW model includes a factor of $e^{-qd}$, reducing the SU(2) symmetry to U(1) for finite layer separation $d$. Consequently, the order parameter of PSDW is significantly smaller than that of SDW.

\begin{figure*}[!t]
\centering
\includegraphics[width=1.0\textwidth]{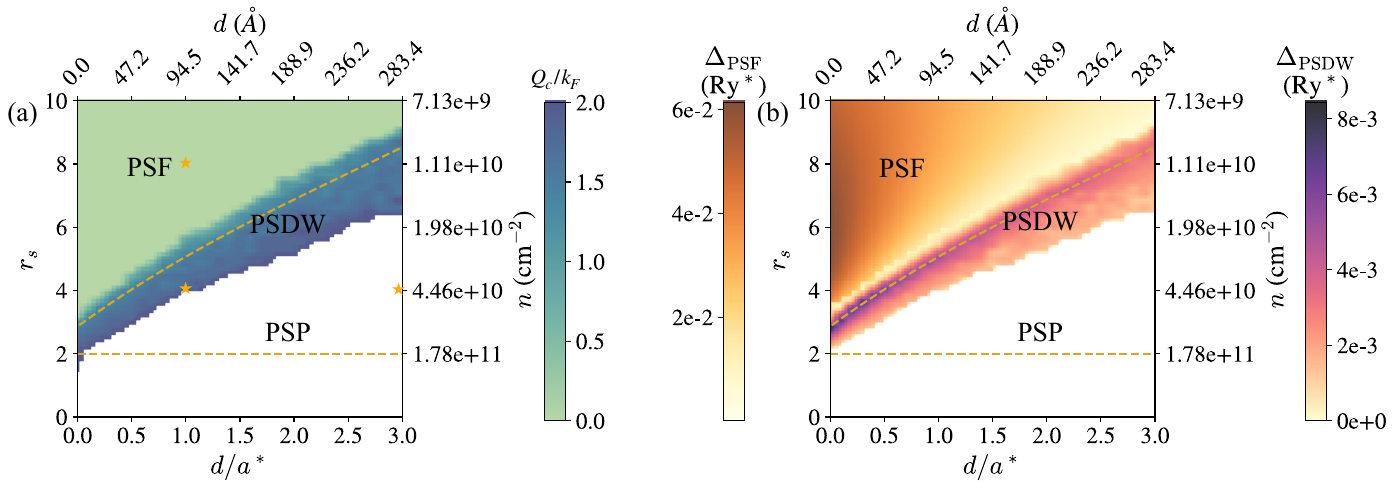}
\caption{\label{fig_PhaseDiagram_2band}
PSDW in 2D electron bilayers. 
(a) The HF phase diagram as a function of ($r_s$,$d$). The HF ground state is homogeneous PSF at large $r_s$ and PSP at large $d$, which agrees with the original phase boundaries calculated by homogeneous HF theory \cite{JZ_interlayerCoherence_2024} as marked by yellow dashed lines.
In the regime between the PSF and PSP phases, the PSDW state has a lower energy across all $r_s$ and $d$ values shown in the phase diagram. This PSDW phase occupies a larger region within the original PSP domain than in the PSF domain, similar to the SDW case in Fig.~\ref{fig_SU2}(a), and expands a broader $r_s$ range as $d$ increases.
The PSDW momentum $Q_c$ is $\sim 2k_F$ near the PSDW-PSP boundary, decreases towards the PSDW-PSF boundary and ultimately drops to zero as approaching the PSF phase.
(b) The stability of PSF and PSDW phases, quantified by the energy difference between the ground state and the second lowest-energy state: in the PSF phase, it is $\Delta_{\rm PSF}=\varepsilon^{\rm PSP}_{\rm tot} - \varepsilon^{\rm PSF}_{\rm tot}$ and in the PSDW phase $\Delta_{\rm PSDW}= \text{min}\{\varepsilon^{\rm PSP}_{\rm tot}, \varepsilon^{\rm PSF}_{\rm tot}\} - \varepsilon^{\rm PSDW}_{\rm tot}$. The maximum of $\Delta_{\rm PSDW}$ is an order of magnitude smaller than that of $\Delta_{\rm PSF}$. The maximum of $\Delta_{\rm PSDW}$ tracks the original PSF-PSP phase boundary, marked by the yellow dashed lines, which is analogous to the observation in SDW that the instability is most pronounced near the paramagnetic to ferromagnetic transition (Fig.~\ref{fig_SU2}(a)). 
Note that the PSDW phase boundaries are interpolated to a denser ($r_s$, $d$) grid and we only consider the PSDW state to have lower energy than PSP or PSF state if the energy difference is greater than $\sim 0.001$ Ry$^*$.
}
\end{figure*}

In Fig.~\ref{fig_PhaseDiagram_2band}(a), we show the HF phase diagram as a function of ($r_s$,$d$) by solving the self-consistent equations from Sec.~\ref{sec_2band}.  The HF ground state transitions from the homogeneous PSF ($Q_c=0$) at large $r_s$ to PSP at large $d$, which agrees with the original phase boundaries calculated by homogeneous HF theory \cite{JZ_interlayerCoherence_2024} as marked by yellow dashed lines.
In the regime between the PSF and PSP phases, a new phase---the PSDW state---that breaks both translational and rotational symmetries is found to have a lower energy across all $r_s$ and $d$ values shown in the phase diagram Fig.~\ref{fig_PhaseDiagram_2band}(a).
This PSDW phase occupies a larger region within the original PSP domain than in the PSF domain, similar to the SDW case shown in Fig.~\ref{fig_SU2}(a).
Additionally, the PSDW phase for a given $d$ expands a broader $r_s$ range as $d$ increases. 
The PSDW momentum $Q_c$, indicated by the color in Fig.~\ref{fig_PhaseDiagram_2band}(a), is approximately $2k_F$ near the PSDW-PSP boundary, decreases towards the PSDW-PSF boundary, and quickly drops to zero as approaching the PSF phase. This behavior of the density wave momentum is analogous to that observed in the Overhauser SDW case depicted in Fig.~\ref{fig_SU2}(b).

The phase diagram is re-presented in Fig.~\ref{fig_PhaseDiagram_2band}(b) by the ground-state stability, characterized by the energy difference between the HF ground state and the second lowest-energy state: in the PSF phase, it is $\Delta_{\rm PSF}=\varepsilon^{\rm PSP}_{\rm tot} - \varepsilon^{\rm PSF}_{\rm tot}$ and in the PSDW phase $\Delta_{\rm PSDW}= \text{min}\{\varepsilon^{\rm PSP}_{\rm tot}, \varepsilon^{\rm PSF}_{\rm tot}\} - \varepsilon^{\rm PSDW}_{\rm tot}$.
As indicated by the color plot in Fig.~\ref{fig_PhaseDiagram_2band}(b), the maximum of $\Delta_{\rm PSDW}$ is an order of magnitude smaller than that of $\Delta_{\rm PSF}$. 
Interestingly, the maximum of $\Delta_{\rm PSDW}$ tracks the original PSF-PSP phase boundary, marked by the yellow dashed line, corroborating the observation in SDW that the instability is most pronounced near the paramagnetic to ferromagnetic transition (Fig.~\ref{fig_SU2}(a)).

\begin{figure*}[!htb]
\centering
\includegraphics[width=1.0\textwidth]{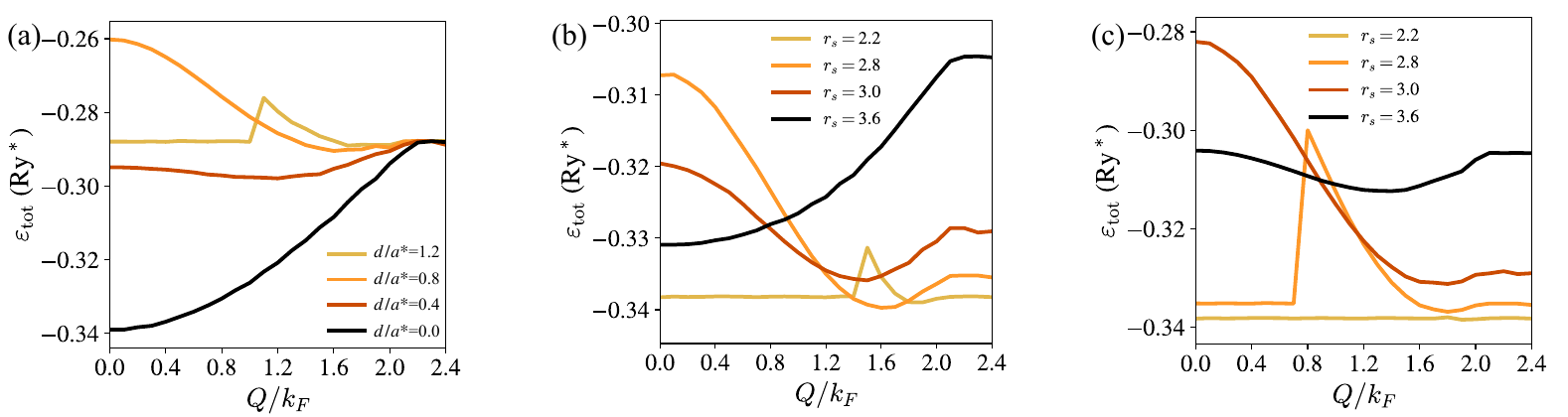}
\caption{\label{fig_eptot_PSDW} {
PSDW in 2D electron bilayers.
The HF energy, $\varepsilon_{\rm tot}$, versus $Q$ for (a) fixed $r_s=4$ and varying $d$,
(b) fixed $d/a^*=0.1$ and varying $r_s$, and (c) fixed $d/a^*=0.3$ and varying $r_s$. 
In (a), as $d$ decreases, the energy of the homogeneous PSF state (at $Q=0$) first rises then falls, while the energy of the PSP state (at $Q \gtrsim 2k_F$) remains unchanged due to the $d$-independence of the interlayer incoherent state. In (b-c), both the PSF and PSP states exhibit changes in energy as $r_s$ varies.
  }}
\end{figure*}

Figure~\ref{fig_eptot_PSDW} illustrates the evolution of the HF energy, $\varepsilon_{\rm tot}$, with respect to $d$ for fixed $r_s=4$ in Fig.~\ref{fig_eptot_PSDW}(a), and with respect to $r_s$ for fixed $d$'s in Fig.~\ref{fig_eptot_PSDW}(b-c). 
As approaching the PSP-PSDW phase boundary from the PSP side, interlayer coherence first occurs in a confined region in k-space near $Q \sim 2k_F$. 
In Fig.~\ref{fig_eptot_PSDW}(a), as $d$ decreases--effectively increasing the interlayer Coulomb interaction--the energy of the homogeneous PSF state (at $Q=0$) first rises then falls, while the energy of the PSP state (at $Q \gtrsim 2k_F$) remains unchanged due to the $d$-independence of the interlayer incoherent state.
In Fig.~\ref{fig_eptot_PSDW}(b-c), both the PSF and PSP states exhibit changes in energy as $r_s$ varies. 
These figures reveal that as the Coulomb interaction strengthens, either by decreasing $d$ or increasing $r_s$, the PSDW state initially overtakes the PSP state, becomes most stable when the PSF and PSP energies equalize, and then gradually loses its energy advantage to the PSF state.
It should be noted that, similar to Figs.~\ref{fig_epxinter_SU2}(b) and \ref{fig_eps_SU2}(a), the sailboat-like $\varepsilon_{\rm tot}$ in Fig.~\ref{fig_eptot_PSDW} (the yellow lines in Fig.~\ref{fig_eptot_PSDW}(a-c) and the orange line in Fig.~\ref{fig_eptot_PSDW}(c)) is a direct result of choosing the PSF state as the HF initial state in these calculations \cite{Unlike}.

\begin{figure*}[!htb]
\centering
\includegraphics[width=1.0\textwidth]{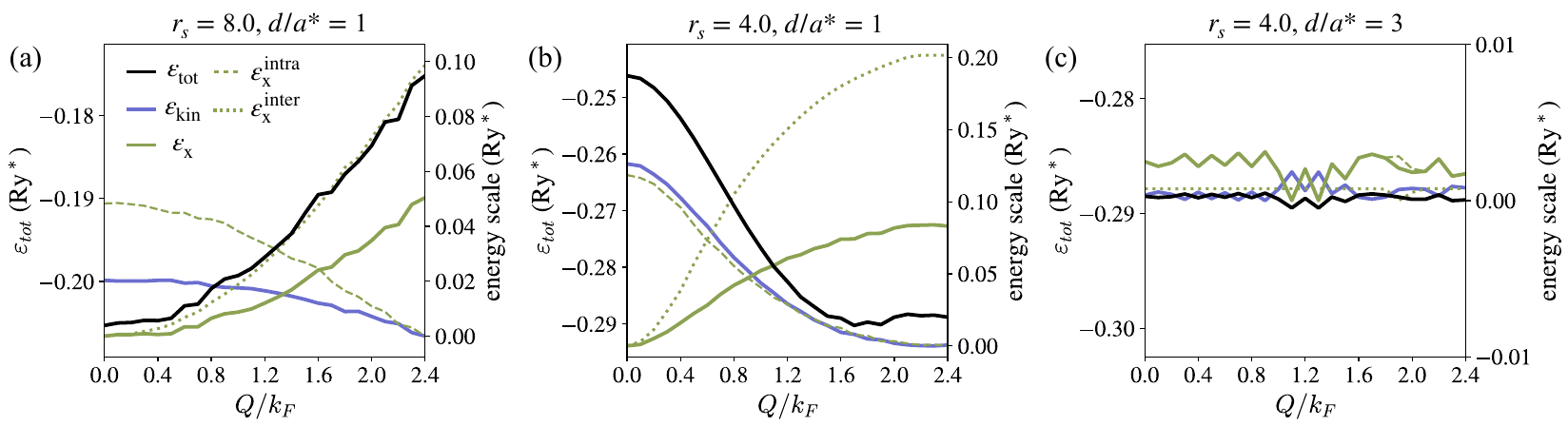}
\caption{\label{fig_eps_PSDW} {
All energy components of PSDW HF energy $\varepsilon_{\rm tot}$ (black solid lines): the kinetic energy $\varepsilon_{\rm kin}$ (blue solid lines) and the exchange energy $\varepsilon_{\rm x}$ (green solid lines), including $\varepsilon_{\rm x}^{\rm intra}$ (green dashed lines) and $\varepsilon_{\rm x}^{\rm inter}$ (green dotted lines). (a) $r_s= 8$, $d/a^*=1$, deep in the PSF phase, (b) $r_s = 4$, $d/a^*=1$, in the PSDW phase and (c) $r_s = 4$, $d/a^*=3$, which is deep in the PSP phase. In (a), the HF energy is minimized at $Q=0$ as $\varepsilon_{\rm x}^{\rm inter}$ dominates and is minimized at $Q=0$.
In (b), $\varepsilon_{\rm tot}$ reaches its minimum at a finite $Q$ as a result of the comparable scale of exchange and kinetic energies.
In (c), there is no obvious energy variation with $Q$. 
In each figure, the $y$-axis on the left measures $\varepsilon_{\rm tot}$, and the $y$-axis on the right scales other energy components ($\varepsilon_{\rm kin}$, $\varepsilon_{\rm x}$, $\varepsilon_{\rm x}^{\rm intra}$ and $\varepsilon_{\rm x}^{\rm inter}$) which are all offset for clarity and comparative convenience in the same figure. 
  }}
\end{figure*}

In Fig.~\ref{fig_eps_PSDW}, we show all energy components as a function of $Q$ at three marked points in the phase diagram Fig.~\ref{fig_PhaseDiagram_2band}(a). These points represent distinct phases: one phase deep in the PSF regime ($r_s=8$, $d/a^*=1$), one phase deep in the PSP regime ($r_s=4$, $d/a^*=3$) and one phase within the PSDW regime ($r_s=4$, $d/a^*=1$).
In the PSF phase (Fig.~\ref{fig_eps_PSDW}(a)), the HF energy is minimized at $Q=0$ as $\varepsilon_{\rm x}^{\rm inter}$ dominates and is minimized at $Q=0$. In the PSP phase (Fig.~\ref{fig_eps_PSDW}(c)), there is no evident energy variation with $Q$. In the PSDW phase (Fig.~\ref{fig_eps_PSDW}(b)), $\varepsilon_{\rm tot}$ reaches its minimum at a finite $Q$ as a result of the comparable scale of exchange and kinetic energies.
Note that in Fig.~\ref{fig_eps_PSDW}(a), unlike the scenario in Fig.~\ref{fig_eps_SU2}, energies do not plateau for large momentum $Q > 2k_F$. This is because with increasing $r_s$, spontaneous coherence remains significant for larger $Q$ values, a phenomenon also observed in the SDW case for larger $r_s$ values (not shown in figures presented in the paper).
The converged HF results, including quasiparticle bands and interlayer coherence order parameter $\Delta_k$ in these three distinct phases are shown in Appendix~\ref{supp_bands_equaln}.

Even though we have focused on the two-band PSDW model, ignoring the spin degree of freedom, there is no obvious reason to rule out the possibility of the spin-pseudospin density wave (S-PSDW), i.e., both the intralayer SDW and interlayer PSDW occur in an e-e bilayer, as schematically shown in Fig.~\ref{fig_schematic}(c) as an example of S-PSDW. Because of the exchange-driven nature of (pseudo)spin density waves, S-PSDW state with the optimized combination of SDW and PSDW momenta, $\mathbf{Q}_1$ and $\mathbf{Q}_2$, should have a lower energy.
However, these two independent density wave momenta complicate the theory, we therefore briefly comment on the four-band model, including both spin and pseudospin degrees of freedom, in  Appendix~\ref{supp_4band}.

\section{The phase diagram of unequal layer densities}
\label{sec_unequal_density}
We extend our analysis of equal layer densities from Sec.~\ref{sec_para_instability} to the case involving unequal layer densities. Our self-consistent HF calculations indicate that the PSDW state is only stable for a small layer density imbalance. Because of the intralayer exchange interaction, all electrons tend to be polarized into one of the layers if either the layer imbalance $m$ or the average inter-electron distance $\bar{r}_s$ is large--the HF theory favors the layer fully polarized state.

In experimental setups, a layer density imbalance is very often induced using a dual-gated structure, in which the chemical potential across the system remains constant. Consequently, a finite displacement field is generated when densities in the two layers are not equal. We model this external displacement field as the electrostatic energy difference, $\varepsilon_g$, between the two 2DEG layers \cite{Under}:
\begin{equation}
\begin{split}
\varepsilon_g 
&= \varepsilon_{F,b} - \varepsilon_{F,t} \\
&= \frac{2\pi\hbar^2}{m^*} n m,
\end{split}
\end{equation}
where $n$ is the total electron density, and $m$ is a dimensionless parameter that is used to tune the strength of the chemical potential difference and induce a density imbalance.
Without loss of generality, we assume $\varepsilon_{F,b} > \varepsilon_{F,t}$, $\varepsilon_g > 0$. 
Under these conditions, the Hamiltonian and the self-consistent equations in Eqs.~(\ref{Eq_Hamil}-\ref{Eq_scEqs_PSDW}) continue to apply, with the addition of terms for the electrostatic energy difference, $\varepsilon_g$, and Hartree energies proportional to the layer density imbalance. Equation~(\ref{Eq_eps}) is modified accordingly to include these factors,
\begin{align}
\varepsilon_{t, \mathbf{k}+\frac{\mathbf{Q}}{2}} &= \frac{\hbar^2|\mathbf{k}+\mathbf{Q}/2|^2}{2m^*} +\varepsilon_g + \frac{2\pi e^2 d}{\epsilon_b} n_t \nonumber\\
& \quad - \frac{1}{A}\sum\limits_{\mathbf{k}'} V^0_{\mathbf{k}'-\mathbf{k}} \rho_{tt}(\mathbf{k}'), \nonumber\\
\varepsilon_{b, \mathbf{k}-\frac{\mathbf{Q}}{2}} &= \frac{\hbar^2|\mathbf{k}-\mathbf{Q}/2|^2}{2m^*} + \frac{2\pi e^2 d}{\epsilon_b} n_b \nonumber\\
& \quad - \frac{1}{A}\sum\limits_{\mathbf{k}'} V^0_{\mathbf{k}'-\mathbf{k}} \rho_{bb}(\mathbf{k}').
\end{align}
The modified self-consistent equations account for these adjustments are
\begin{align}
\xi_{\mathbf{k}} &= \frac{\hbar^2}{4m^*} \Big( |\mathbf{k}+\frac{\mathbf{Q}}{2} |^2 - |\mathbf{k}-\frac{\mathbf{Q}}{2} |^2 \Big) \nonumber\\
& \quad + \frac{\varepsilon_g}{2} + \frac{\pi e^2 d}{\epsilon_b}(n_t-n_b) \\
& \quad + \frac{1}{A} \sum\limits_{\mathbf{k}'} V^0_{\mathbf{k}'-\mathbf{k}} \frac{\xi_{\mathbf{k}'}}{\sqrt{\xi_{\mathbf{k}'}^2 + \Delta_{\mathbf{k}'}^2}} (f_{-,\mathbf{k}'} - f_{+,\mathbf{k}'}), \nonumber\\
\Delta_{\mathbf{k}} &= \frac{1}{2A} \sum\limits_{\mathbf{k}'} V^d_{\mathbf{k}'-\mathbf{k}} \frac{\Delta_{\mathbf{k}'}}{\sqrt{\xi_{\mathbf{k}'}^2 + \Delta_{\mathbf{k}'}^2}} (f_{-,\mathbf{k}'} - f_{+,\mathbf{k}'}). \nonumber
\end{align}

\begin{figure*}[!t]
\centering
\includegraphics[width=0.9\textwidth]{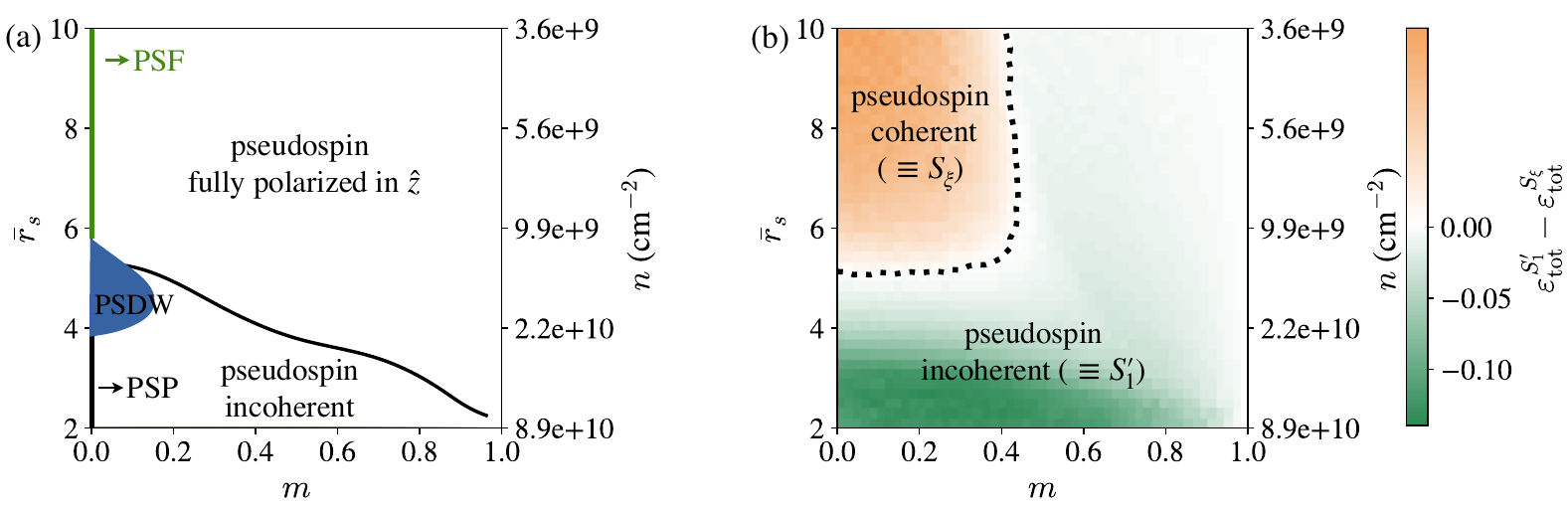}
\caption{\label{fig_PhaseDiagram_unequal}
PSDW in 2D electron bilayers.
Phase diagram of unequal layer densities, with a fixed $d/a^*=1$, as a function of ($m$,$\bar{r}_s$), where $\bar{r}_s = (2/\pi n)^{1/2}/a^*$ is the average inter-electron distance. (a) The schematic phase diagram of self-consistent HF calculations. In the $m=0$ limit, i.e., equal layer density case, the diagram recovers the phases found at $d/a^*=1$ in Fig.~\ref{fig_PhaseDiagram_2band}(a): PSP for $\bar{r}_s \lesssim 4$, PSDW for $4 \lesssim \bar{r}_s \lesssim 6$, and PSF for $\bar{r}_s \gtrsim 6$. 
When $m$ is nonzero, PSDW can in principle only occur at small layer density imbalances. As $m$ increases, the energy benefit from the PSDW formation is exponentially small and the interlayer coherence is confined to a tiny region in momentum space as a perturbation to the paramagnetic state.
At larger $m$ or $\bar{r}_s$ values, the dominant intralayer exchange energy leads to a full layer polarization, i.e., the pseudospin fully polarized phase with an Ising order or equivalently the $S_3$ phase in Ref.~\cite{JZ_interlayerCoherence_2024}). At smaller $\bar{r}_s$, pseudospins are incoherent.
(b) The phase diagram by directly comparing the HF energies between the pseudospin coherent and pseudospin incoherent states, denoted as $S_\xi$ and $S_1'$ in Ref.~\cite{JZ_interlayerCoherence_2024}, respectively. The HF energies are calculated using the same $k$-space grid as in (a).
}
\end{figure*}

The HF energy, $\varepsilon_{\rm tot}$, now incorporates additional electrostatic energy and Hartree energy reflective of the layer density imbalance,
\begin{align}
\label{Eq_E_imbalance}
\varepsilon_{\rm tot} &= \varepsilon_{\rm kin} + \varepsilon_{\rm H} + \varepsilon_{\rm x}^{\rm intra} + \varepsilon_{\rm x}^{\rm inter}, \nonumber\\
\varepsilon_{\rm kin} &= \frac{A n_{t} \varepsilon_g}{N} + \frac{\hbar^2}{2m^*N} \sum\limits_{\mathbf{k}} \big|\mathbf{k}
+ \frac{\mathbf{Q}}{2}\big|^2 \rho_{tt}(\mathbf{k}) \nonumber\\
& \qquad \qquad \ + \frac{\hbar^2}{2m^*N} \sum\limits_{\mathbf{k}} 
 \big|\mathbf{k} - \frac{\mathbf{Q}}{2} \big|^2 \rho_{bb}(\mathbf{k}) , \nonumber\\
\varepsilon_{\rm H} &= \frac{\pi e^2 A d}{2 \epsilon_b N} (n_t-n_b)^2, \\
\varepsilon_{\rm x}^{\rm intra} &= -\frac{1}{2AN} \sum\limits_{\mathbf{k}, \mathbf{k}'} V^0_{\mathbf{k}'-\mathbf{k}} \big[ \rho_{tt}(\mathbf{k}') \rho_{tt}(\mathbf{k}) \nonumber\\
& \qquad \qquad \qquad \qquad + \rho_{bb}(\mathbf{k}') \rho_{bb}(\mathbf{k}) \big],
\nonumber\\
\varepsilon_{\rm x}^{\rm inter} 
&= -\frac{1}{AN} \sum\limits_{\mathbf{k}, \mathbf{k}'} V^d_{\mathbf{k}'-\mathbf{k}}
\rho_{tb}(\mathbf{k}') \rho^*_{tb}(\mathbf{k}). \nonumber
\end{align}

It should be noted that, the electrostatic energies in Eq.~(\ref{Eq_E_imbalance}) specifically depend on the dual-gated experimental setup we described previously. 
The self-consistent HF phase diagram, with a fixed $d/a^*=1$, is schematically shown in Fig.~\ref{fig_PhaseDiagram_unequal}(a) as a function of ($m, \bar{r}_s$). Here, $\bar{r}_s = (2/\pi n)^{1/2}/a^*$ is the average inter-electron distance. In the $m=0$ limit, which corresponds to equal layer densities, the diagram reflects the phases found at $d/a^*=1$ shown in Fig.~\ref{fig_PhaseDiagram_2band}(a): PSP for $\bar{r}_s \lesssim 4$, PSDW for $4 \lesssim \bar{r}_s \lesssim 6$, and PSF for $\bar{r}_s \gtrsim 6$.
When $m$ is nonzero, PSDW can in principle only occur at small layer density imbalances. 
As $m$ increases, the energy benefit from the PSDW formation is exponentially small and the interlayer coherence is confined to a tiny region in momentum space as a perturbation to the paramagnetic state.
At larger $m$ or $\bar{r}_s$ values, the intralayer exchange energy becomes dominant, leading to a full polarization of electrons into one layer, which is the pseudospin fully polarized phase with an Ising order (in $\hat{z}$-direction, i.e., $S_3$ phase in Ref.~\cite{JZ_interlayerCoherence_2024}). At smaller $\bar{r}_s$, pseudospins are incoherent, resulting in uneven electron distribution across the two layers.

Figure~\ref{fig_PhaseDiagram_unequal}(b) shows the phase diagram by directly comparing the HF energies between the pseudospin coherent state and pseudospin incoherent state, denoted as $S_\xi$ and $S_1'$ \cite{Compared} in Ref.~\cite{JZ_interlayerCoherence_2024}, respectively. Expressed in the summation of $k$-space occupied states,
\begin{align}
\label{Eq_eptot_k_unequal}
\varepsilon_{\rm tot}^{S_{\xi}} &= \frac{An_t \varepsilon_g}{N} + \frac{\hbar^2}{2m^*N} \sum\limits_{k \leq k^{S_\xi}_F} k^2 + \frac{\pi e^2Ad}{2\epsilon_b N}(n_t - n_b)^2 \nonumber \\
& \quad - \frac{\pi e^2}{\epsilon_b NA} \sum\limits_{\substack{k,k' \\ \leq k^{S_\xi}_F}
} \frac{\alpha^4 + \beta^4 + 2 \alpha^2\beta^2 e^{-|\mathbf{k} - \mathbf{k}'|d}}{|\mathbf{k} - \mathbf{k}'|}, \\
\varepsilon_{\rm tot}^{S_1'} &= \frac{An_t \varepsilon_g}{N} + \frac{\hbar^2}{2m^*N} \Big(\sum\limits_{k \leq k_{F,t}} k^2 \alpha^2 + \sum\limits_{k \leq k_{F,b}} k^2 \beta^2 \Big) \nonumber\\ 
& \quad - \frac{\pi e^2}{\epsilon_b NA} \Bigg(\sum\limits_{\substack{k,k' \\ \leq k_{F,t}} }\frac{\alpha^4}{|\mathbf{k} - \mathbf{k}'|} + \sum\limits_{\substack{k,k' \\ \leq k_{F,b}}} \frac{\beta^4}{|\mathbf{k} - \mathbf{k}'|} \Bigg),
\end{align}
where $n_t = n\alpha^2$, $n_b = n\beta^2$, and $\alpha^2+\beta^2=1$. $k_F^{S_\xi} = (4\pi n)^{1/2}$ is the Fermi momentum of $S_\xi$ state, $k_{F,t} = (4\pi n_t)^{1/2}$ and $k_{F,b} = (4\pi n_b)^{1/2}$ are Fermi momenta in $S_1'$ state. 
Figure~\ref{fig_PhaseDiagram_unequal}(b) is calculated using the same $k$-space grid as in Fig.~\ref{fig_PhaseDiagram_unequal}(a). 
The difference from Fig.~3(f) in Ref.~\cite{JZ_interlayerCoherence_2024} in the phase diagram Fig.~\ref{fig_PhaseDiagram_unequal}(b) is attributed to the consideration of additional electrostatic energy, which weakens the exchange energy benefits in the $S_\xi$ state.

Comparing Figs.~\ref{fig_PhaseDiagram_unequal}(a) and (b), PSDW tends to occur near the $S_\xi$-$S_1'$ boundary, similar to the behavior seen in the equal layer density case in Fig.~\ref{fig_PhaseDiagram_2band}. The boundary of pseudospin incoherent phase and pseudospin fully polarized phase with an Ising order in Fig.~\ref{fig_PhaseDiagram_unequal}(a) qualitatively aligns with the zero energy-difference in Fig.~\ref{fig_PhaseDiagram_unequal}(b).

\section{RPA static screening}
\label{sec_RPA}
Just as correlation effects suppress ferromagnetic instability, they also undermine the stability of SDW or PSDW \cite{Perdew_1980, Brener_1981}.
It is well established that screening effects always favor the paramagnetic state \cite{Fedders_correlation_1966, Amusia_SDW_screen_1966}.
Previous studies have demonstrated the fragility of the Overhauser SDW instability to correlation effects \cite{Wolff_RPA_1960, Fedders_correlation_1966, Hamann_correlation_1966}.
In a dense electron gas, any perturbation to the paramagnetic state becomes local and such local instabilities particularly susceptible to Thomas-Fermi screening effect.
Our HF calculations in Sec.~\ref{sec_para_instability}-\ref{sec_unequal_density} show that at high densities, both inter-spin and interlayer coherence become local (with large $Q_c$), and the corresponding exchange energy gain from the formation of SDW or PSDW is insufficient to counterbalance the impact of correlations \cite{Giuliani_SDW_2008}.
In this section, we explore the HF phase diagram by self-consistently taking into account the RPA static screening effect and further correcting the model with Hubbard-type local field corrections.
We find that RPA static screening eliminates all coherent phases, leaving only the spin and pseudospin paramagnetic phase in the phase diagram. When Hubbard-type local field corrections are included, spin ferromagnetic and PSF phases re-emerge, but SDW and PSDW remain absent. Despite these results, a rigorous treatment of dynamical screening theory could potentially rescue SDW and PSDW, as unscreened (static RPA) are approximations over- (under-) estimating the importance of screening, and the dynamically screened theories typically give results in between, closer to the unscreened approximation \cite{Sreejith_2024, SDS_screen_1990}.
However, developing such a rigorous and appropriate RPA dynamical screening theory is beyond the scope of the current work.

In RPA static screening theory, the Coulomb interaction $V^d_\mathbf{q}$ is screened by the static dielectric constant by
\begin{equation}
\label{Eq_Vsc}
\begin{split}
V^d_{\rm sc}(\mathbf{q}) = \epsilon^{-1}(\mathbf{q}) V^d(\mathbf{q})
= \epsilon^{-1}(\mathbf{q}) \frac{2\pi e^2}{\epsilon_{\rm b} q} e^{-qd},
\end{split}
\end{equation}
where $\epsilon_{\rm b}$ is the dielectric constant of the surrounding medium.
The static dielectric function is
\begin{equation}
\label{Eq_epsilon_RPA}
\begin{split}
\epsilon(\mathbf{q}) = 1-V^d(\mathbf{q}) \chi_0(\mathbf{q}),
\end{split}
\end{equation}
where $\chi_0(\mathbf{q})$ is the static polarization function. For a paramagnet, using the Lindhard formula,
\begin{equation}
\label{Eq_chi}
\begin{split}
\chi_0(\mathbf{q}) = \frac{g}{A} \sum\limits_{n,m,\mathbf{k}} \frac{f_{n \mathbf{k}} - f_{m \mathbf{k}+\mathbf{q}}}{\varepsilon_{n \mathbf{k}} - \varepsilon_{m \mathbf{k}+\mathbf{q}}} 
\big| \langle m \mathbf{k}+\mathbf{q} | e^{i\mathbf{q} \cdot \mathbf{r}} | n \mathbf{k} \rangle \big|^2,
\end{split}
\end{equation}
where $g$ is the spin or pseudospin degeneracy, $n$ and $m$ label quasiparticle bands, and $|n \mathbf{k} \rangle$ is the quasiparticle eigenvector.
In the long wavelength limit, $q \rightarrow 0$, the intraband transitions contribute a constant to the polarization function, which is just the density of states at the Fermi level $\mathcal{D}(\varepsilon_{\rm F})$,
\begin{equation}
\begin{split}
\lim\limits_{q \rightarrow 0} \chi^{\rm intra}_0(\mathbf{q}) 
&= \lim\limits_{q \rightarrow 0} \frac{1}{A} \sum\limits_{n,\mathbf{k}} \frac{f_{n \mathbf{k}} - f_{n \mathbf{k}+\mathbf{q}}}{\varepsilon_{n \mathbf{k}} - \varepsilon_{n \mathbf{k}+\mathbf{q}}} \\
&= - \mathcal{D}(\varepsilon_{\rm F}).
\end{split}
\end{equation}
The screened Coulomb interaction in the long wavelength limit is the inverse of the density of states,
\begin{equation}
\begin{split}
\lim\limits_{q \rightarrow 0} V_{\rm sc}(\mathbf{q}) = \mathcal{D}^{-1}(\varepsilon_{\rm F}).
\end{split}
\end{equation}

\begin{figure}[!t]
\centering
\includegraphics[width=0.8\columnwidth]{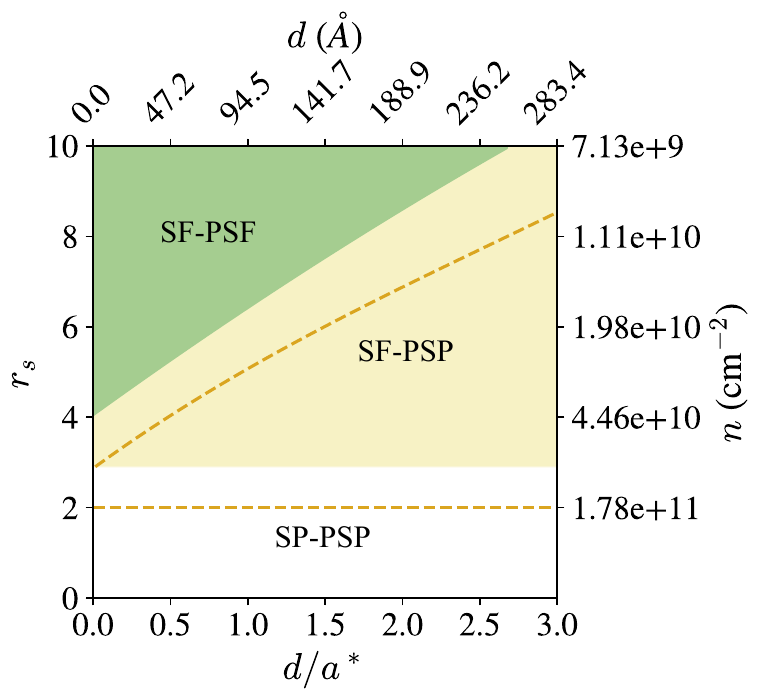}
\caption{\label{fig_PhaseDiagram_2band_wscreen}
{The RPA screened self-consistent HF phase diagram of equal layer densities, with paramagnetic local field correction in Eq.~(\ref{Eq_epsilon_para}).
At small densities (large $r_s$) and small $d$, lowest energy state is spin and pseudospin ferromagnetic (SF-PSF). For intermediate $r_s$ and large $d$, it is the spin ferromagnetic and pseudospin paramagnetic (SF-PSP) state. For $r_s \lesssim 3$, it is the spin and pseudospin paramagnetic (SP-PSP) state.
The yellow dashed lines trace the HF phase boundary without screening effects \cite{JZ_interlayerCoherence_2024}.
}}
\end{figure}

RPA static screening stabilizes both spin and pseudospin paramagnetic states until extremely high $r_s$ values, leading to the disappearance of spin and pseudospin ferromagnetic, SDW, and PSDW states from the phase diagram.
However, the most primitive RPA static screening theory tends to overestimate the strength of screening since it only takes into account the Hartree part. 
To incorporate exchange corrections to the RPA approximation, we further include Hubbard-type local field factors $G_{ss'} = G_{ss} \delta_{ss'}$, which ignores correlations between opposite (pseudo)spins, $G_{s\bar{s}}$.
For a general case with any (pseudo)spin polarization,
\begin{equation}
\begin{split}
G_{ss}(q) &= \frac{q}{\sqrt{q^2 + k^2_{F,s}}},
\end{split}
\end{equation}
The static dielectric constant for (pseudo)spin $s$ (Appendix~\ref{Appendix_RPA}) is \cite{DJKim_negative_epsilon_1972}
\begin{equation}
\label{Eq_epsilon_ferro}
\begin{split}
\epsilon_s(\mathbf{q}) = (1 + V^d(\mathbf{q})G_{s s} \chi_{0s}) \big[ 1-V^d(\mathbf{q})\sum\limits_{s'}\tilde{\chi}_{s'} \big],
\end{split}
\end{equation}
where $\tilde{\chi}_s$ is the proper density response function including the effect of exchange interactions and is related to the Lindhard function $\chi_{0s}$ by
\begin{equation}
\begin{split}
\tilde{\chi}_s^{-1} = \chi_{0s}^{-1} + V^d(q) G_{s s}.
\end{split}
\end{equation}
In the paramagnetic state, $\chi_{0s} = \chi_{0\bar{s}} = \chi_0/2$, Eq.~(\ref{Eq_epsilon_ferro}) reduces to the dielectric constant of original Hubbard's result for a paramagnet:
\begin{equation}
\label{Eq_epsilon_para}
\begin{split}
&G_{ss} = G_{\bar{s}\bar{s}} = G, \\
\epsilon_{s} = \epsilon_{\bar{s}}
&= 1- V^d(q) (1-\frac{G}{2}) \chi_0.
\end{split}
\end{equation}

The screened HF phase diagram is calculated by self-consistently evaluating the screened Coulomb potential Eq.~(\ref{Eq_Vsc}), accounting for, from the most screened to the least screened, the primitive RPA static screening, RPA with ferromagnetic and paramagnetic local field factor corrections.
The primitive RPA static screening eliminates all coherent phases, including spin ferromagnetic and SDW phases in the single-layer 2DEG (Appendix~\ref{Appendix_RPA_spin}), as well as PSF and PSDW phases, across all considered parameters ($r_s$,$d$). The primitive RPA static screening leaves only the spin and pseudospin paramagnetic ($S_0$ in Ref.\cite{JZ_interlayerCoherence_2024}) phase in the phase diagram.
The inclusion of Hubbard-type local field corrections reintroduces the spin ferromagnetic and PSF phases into the phase diagram, with phase boundaries shifted to higher $r_s$ values due to correlations. Both SDW and PSDW are fragile to correlations.
Figure~\ref{fig_PhaseDiagram_2band_wscreen} shows the RPA statically screened HF phase diagram for equal layer densities, with corrections from the paramagnetic local field factor Eq.~(\ref{Eq_epsilon_para}). The yellow dashed lines trace the paramagnetic to ferromagnetic phase transition boundaries in the absence of screening effects \cite{JZ_interlayerCoherence_2024}.

In Appendix~\ref{Appendix_RPA_spin}, Fig.~\ref{fig_S1_RPA} shows the critical temperature $T_c$ of the spin ferromagnetic state in a single-layer 2DEG, calculated using the finite-temperature self-consistent HF with RPA static screenings. 
With the inclusion of static screenings, ranging from the weakest (RPA with paramagnetic local field factor correction) to the strongest (the primitive RPA without local field factor corrections) effect, correlations increasingly suppress
$T_c$ and elevate the critical $r_s$. In our calculations, considering a maximum $r_s=10$, spin is not ordered
under the primitive RPA static screening theory.

\section{Discussion}
\label{sec_discussion}
In this paper, we investigate the PSDW instability in 2D electron bilayers, providing the phase diagram based on self-consistent HF theory. 
For bilayers with equal layer densities, we find that the PSDW, characterized by the momentum $Q_c \sim 2k_F$, has the lowest energy near the PSP-PSF phase transition boundary across all layer separations $d$ and electron densities characterized by $r_s$.
For unequal layer densities, the stability of PSDW decreases quickly with increasing layer density imbalance.
Our self-consistent HF calculation is further supplemented by RPA static screening, which eliminates all coherent phases (spin ferromagnetic, SDW, PSF and PSDW). However, after adjusting for Hubbard-type local field corrections, the spin ferromagnetic and PSF phases re-emerge, even though SDW and PSDW are still absent in the phase diagram.
Experimental detection strategies for PSDW, as discussed in Sec.~\ref{sec_intro}, include magneto-transport experiments capable of elucidating Fermi surface properties, anomalies in zero-field conductivity and ARPES.
In bilayer systems, interlayer tunneling is often unavoidable when $d$ is small. 
This tunneling suppresses the PSDW order, similar to the suppression in homogeneous interlayer coherence \cite{JZ_interlayerCoherence_2024}.

We mention that the mean field HF theory, while being qualitatively correct, often quantitatively overestimates various symmetry breaking instabilities, which is why we also provide the results for the screened HF theories in this work.  Typically, an instability or phase transition predicted within the HF theory occurs at stronger interaction compared with the critical interaction strength predicted in theories beyond the mean-field level.  In the current problem involving Coulomb instabilities in e-e bilayers, the dimensionless interaction is characterized by $r_s$ and $d/a^*$, and we expect the experimental critical values for $r_s$ ($d/a^*$) to be higher (lower) than those predicted by our unscreened Hartree-Fock theory, as obtained in our screened mean-field theories.  The important point to emphasize, however, is that both of these parameters, $r_s$ and $d/a^*$, can be continuously varied in 2D bilayers, thus enabling an experimental approach to the interesting PSDWs predicted in our work. (This could not be done for the 3D Overhauser instability where metallic carrier densities cannot be varied by much.)  Since the predicted PSDW is completely novel ground state never before considered in the literature, we hope that our theory will lead to experiments on various electron bilayers at low carrier densities (i.e., large $r_s$) and small layer separations to look for this interesting quantum phase of matter.

Since the observation of spontaneous CDWs and SDWs originally proposed by Overhauser in single-layer 2DEGs is challenging, the same applies to layer PSDWs in 2DEG bilayers.
However, 2D systems with complex band structures and Fermi surfaces show promise for realizing PSDWs.
In general, the pseudospin concept \cite{FZhang_pseudospin_2011}, which is closely analogous to real spins, applies to any two-level system.
This includes the layer degree of freedom in bilayers (as discussed in this paper), sub-bands in wide quantum wells, sublattice or valley degrees of freedom in honeycomb and triangular lattice materials, and cyclotron orbits in Landau levels \cite{Giuliani_SDW_LL_1985, Giuliani_SDW_LL_1986, Giuliani_exciton_LL_1985, Piazza_SDW_LL_1999, Marinescu_SDW_2000, Yoshizawa_SDW_2007, Yoshizawa_SDW_QH_2009, Zheng_SDW_2011}. 
In multilayer graphene moiré systems, the intervalley coherent state has been proposed as the ground state at charge neutrality \cite{Bultinck_hidden_2020, Po_IVC_2018}, and Kekul\'e spiral order at nonzero integer fillings \cite{YHKwan_kekule_2021}.
In these systems, two valleys in momentum space are approximately decoupled due to large momentum separation, and thus are analogous to the layer degree of freedom in nearly decoupled bilayers.
Even in graphene multilayer systems without moiré superlattices, the intervalley coherent phase can occupy a large portion of the phase diagram \cite{Chatterjee_IVC_2022, MXie_SOC_BLG_2023, CHuang_QuarterMetal_2023, DCLu_RTG_2022, CHuang_RTG_2023} and is closely related to unconventional superconductivity.
We anticipate that general PSDWs, involving pseudospin in the sense of layer or valley, are likely to occur in a bilayer structure composed of two rhombohedral multilayer graphenes (RMGs) separated by a thin dielectric film, which prevents tunneling between the two RMGs while allowing interactions.
By applying dual-gated and/or bias voltages, the displacement field between RMGs and the Fermi level in each RMG can be individually tuned. 
The trigonal warping in the band structure, resulting from next-nearest-neighbor hoppings in RMGs, can facilitate the formation of PSDWs by increasing the number of available states for coherence near the Fermi surface.
Although our current study, based on the 2DEG model with circular Fermi surfaces, may not directly relate to experiments, it provides a foundational framework for future investigations into novel coherent behaviors in general 2D bilayers under varying electronic and structural conditions. Detailed exploration of these phenomena is left for future work.

\section{Acknowledgments}
This work is supported by the Laboratory for Physical Sciences. T. C. is supported by a University of California Presidential Postdoctoral Fellowship and acknowledges support from the Gordon and Betty Moore Foundation through Grant No. GBMF8690 to UC Santa Barbara. This research was supported in part by Grant No. NSF PHY-2309135 to the Kavli Institute for Theoretical Physics (KITP). Use was made of computational facilities purchased with funds from the National Science Foundation (CNS-1725797) and administered by the Center for Scientific Computing (CSC). The CSC is supported by the California NanoSystems Institute and the Materials Research Science and Engineering Center(MRSEC; NSF DMR 2308708) at UC Santa Barbara.

\appendix
\section{HF converged quasiparticle bands and interlayer coherence order parameters}
\label{supp_bands_equaln}
\begin{figure*}[!htb]
\centering
\includegraphics[width=1.0\textwidth]{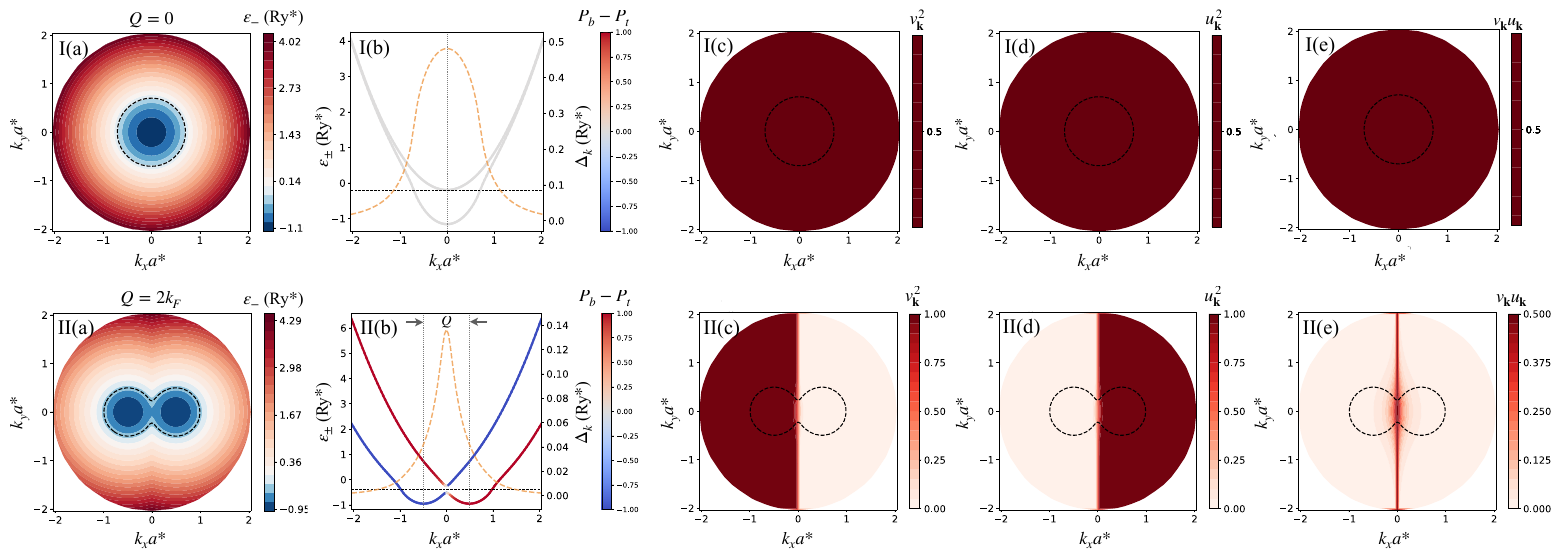}
\caption{\label{fig_rs4_d1} {
Converged HF results in the PSDW phase, $r_s=4$, $d/a^*=1$.
I(a) and II(a) show $k$-space distributions of $\varepsilon_{-,\mathbf{k}}$, in which the black dashed lines plot the Fermi surface contours. 
I(b) and II(b) show quasiparticle bands $\varepsilon_{\pm, k}$ (solid lines) and the interlayer coherence order parameter $\Delta_k$ (yellow dashed lines). The black horizontal dashed lines are Fermi levels. The color of solid lines represents the layer polarization, $P_b - P_t$. 
I(c-e) and II(c-e) show quasiparticle wavefunction probabilities $v_{\mathbf{k}}^2$, $u_{\mathbf{k}}^2$, and $v_{\mathbf{k}}u_{\mathbf{k}}^*$ defined in Eqs.~(\ref{Eq_eigens}-\ref{Eq_expectations}). Panels I(a-e) are for $Q=0$, panels II(a-e) are for $Q=2k_F$. 
  }}
\end{figure*}

\begin{figure*}[!htb]
\centering
\includegraphics[width=1.0\textwidth]{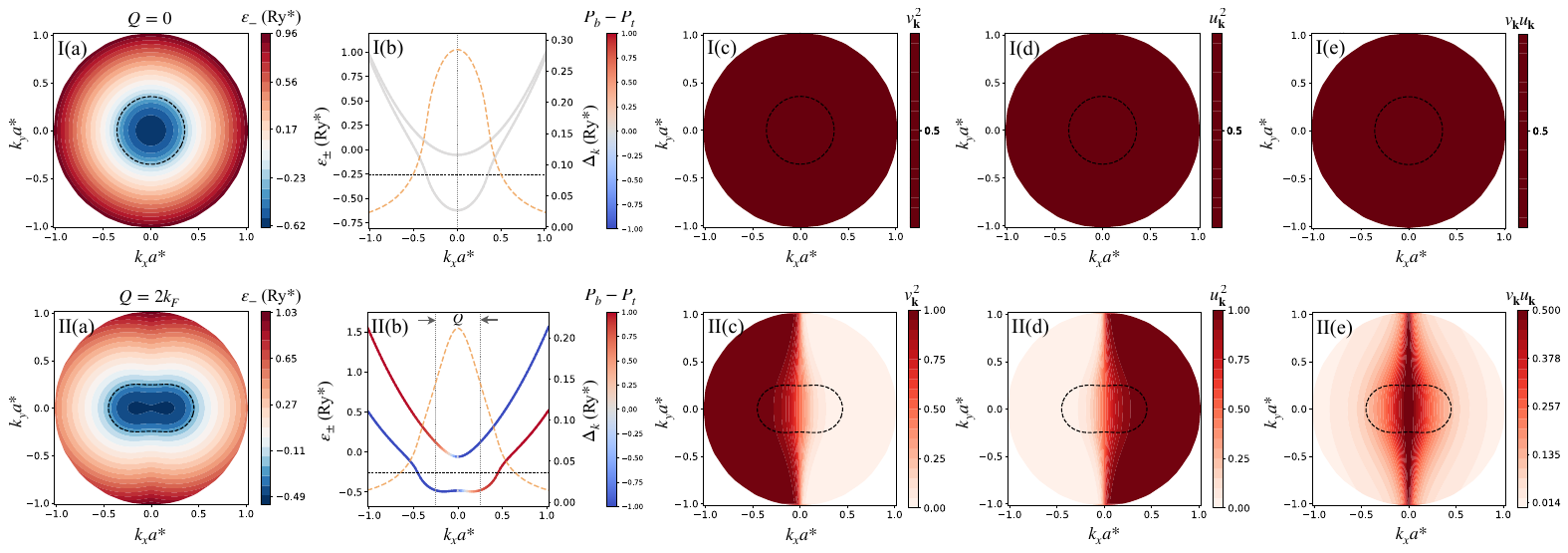}
\caption{\label{fig_rs8_d1} {
Converged HF results in the PSF phase, $r_s=8$, $d/a^*=1$. Same as Fig.~\ref{fig_rs4_d1}.
  }}
\end{figure*}

\begin{figure*}[!htb]
\centering
\includegraphics[width=1.0\textwidth]{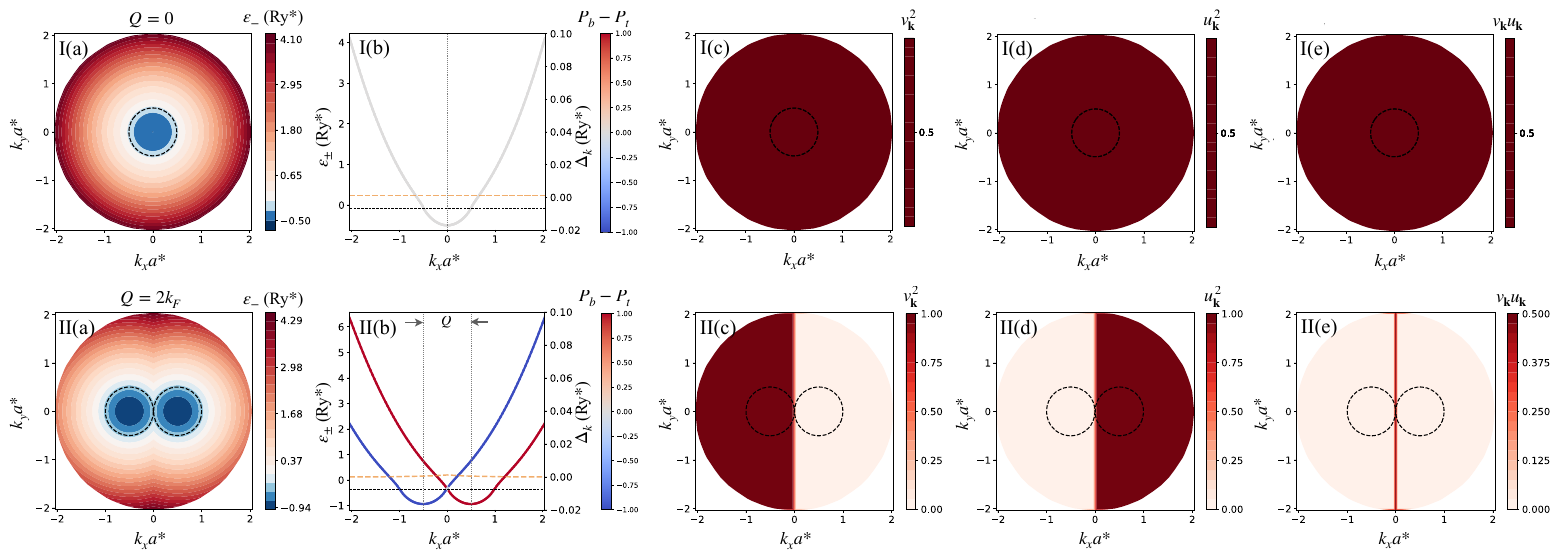}
\caption{\label{fig_rs4_d3} {
Converged HF results in the PSP phase, $r_s=4$, $d/a^*=3$. Same as Fig.~\ref{fig_rs4_d1}.
  }}
\end{figure*}

Figures~\ref{fig_rs4_d1}-\ref{fig_rs4_d3} show the converged HF results in three distinct phases for equal layer densities: one phase within the PSDW regime (Fig.~\ref{fig_rs4_d1}, $r_s=4$, $d/a^*=1$), one phase deep in the PSF regime (Fig.~\ref{fig_rs8_d1}, $r_s=8$, $d/a^*=1$) and one phase deep in the PSP regime (Fig.~\ref{fig_rs4_d3}, $r_s=4$, $d/a^*=3$).

\section{Spin-pseudospin density wave (S-PSDW)}
\label{supp_4band}
In e-e bilayers, the HF ground state for $r_s \lesssim 2$ is spin and pseudospin paramagnetic ($S_0$ phase in Ref.~\cite{JZ_interlayerCoherence_2024}). Therefore, the spin and pseudospin degrees of freedom should be treated on an equal footing, and the SDW and PSDW can in principle both occur in an e-e bilayer. An example of the S-PSDW phase is schematically shown in Fig.~\ref{fig_schematic}(c). 

We present a special model of S-PSDW below, restricting that SDW and PSDW having the same momentum, in which primitive signatures of non-vanishing intralayer SDW and interlayer PSDW order parameters are seen in our self-consistent HF calculations. However, independent density wave momenta make it challenging to find the optimal SDW and PSDW momenta combination which lower the total energy.

With the four-component spinor basis $(c_{t\downarrow,\mathbf{k}+\mathbf{Q}/2}, c_{t\uparrow,\mathbf{k}-\mathbf{Q}/2}, c_{b\downarrow,\mathbf{k}+\mathbf{Q}/2}, c_{b\uparrow,\mathbf{k}-\mathbf{Q}/2})^T$, the HF Hamiltonian in Eq.~(\ref{eq_hamil_2band}) is generalized to 
\begin{widetext}
\begin{equation}
\label{eq_hamil_4band}
\begin{split}
H(\mathbf{k}) = 
\begin{pmatrix}
\varepsilon_{t\downarrow, \mathbf{k}+\mathbf{Q}/2} & -g_{t,\mathbf{k}}(\mathbf{Q}) & -\Delta_\mathbf{k}(0) & -\Delta_\mathbf{k}(\mathbf{Q}) \\
-g^*_{t,\mathbf{k}}(\mathbf{Q}) & \varepsilon_{t\uparrow, \mathbf{k}-\mathbf{Q}/2} & -\Delta_\mathbf{k}(-\mathbf{Q}) & -\Delta_\mathbf{k}(0) \\
-\Delta^{*}_\mathbf{k}(0) & -\Delta^{*}_\mathbf{k}(-\mathbf{Q}) & \varepsilon_{b\downarrow, \mathbf{k}+\mathbf{Q}/2} & -g_{b,\mathbf{k}}(\mathbf{Q}) \\
-\Delta^{*}_\mathbf{k}(\mathbf{Q}) & -\Delta^{*}_\mathbf{k}(0) & -g^*_{b,\mathbf{k}}(\mathbf{Q}) & \varepsilon_{b\uparrow, \mathbf{k}-\mathbf{Q}/2} 
\end{pmatrix}.
\end{split}
\end{equation}
\end{widetext}
Since we only focus on the possibility of the spontaneous SDW and PSDW, but not a rigorous phase diagram, we have assumed that the intralayer SDW and interlayer PSDW have the same momentum $\mathbf{Q}$. 
Note that if the momenta are generically different, the Hamiltonian does not admit such a simple form as Eq.~\eqref{eq_hamil_4band} and the necessary matrix to diagonalize will grow with system size.
The diagonal matrix elements of $H(\mathbf{k})$ are
\begin{equation}
\begin{split}
\varepsilon_{l \sigma, \mathbf{k} \pm \mathbf{Q}/2} &= \frac{\hbar^2|\mathbf{k} \pm \mathbf{Q}/{2}|^2}{2m^*} \\
&\quad - \frac{1}{A}\sum\limits_{\mathbf{k}'} V^0_{\mathbf{k}'-\mathbf{k}} \langle c^\dagger_{l \sigma,\mathbf{k}' \pm \frac{\mathbf{Q}}{2}} c_{l \sigma,\mathbf{k}' \pm \frac{\mathbf{Q}}{2}} \rangle, \\
\end{split}
\end{equation}
$l=t,b$ label layer and $\sigma = \uparrow, \downarrow$ label spin degrees of freedom.
The intralayer SDW order parameters are
\begin{equation}
\label{Eq_g_SPSDW}
\begin{split}
g_{l,\mathbf{k}}(\mathbf{Q}) = \frac{1}{A} \sum\limits_{\mathbf{k}'} V^0_{\mathbf{k}'-\mathbf{k}} \langle c^\dagger_{l \uparrow, \mathbf{k}'-\frac{\mathbf{Q}}{2}} c_{l \downarrow, \mathbf{k}'+\frac{\mathbf{Q}}{2}} \rangle.
\end{split}
\end{equation}
The homogeneous interlayer coherence order parameter $\Delta_\mathbf{k}(0)$ and the PSDW order parameter $\Delta_\mathbf{k}(\pm \mathbf{Q})$ are, respectively,
\begin{equation}
\label{Eq_Delta_SPSDW}
\begin{split}
\Delta_\mathbf{k}(0)
&= \frac{1}{A}\sum\limits_{\mathbf{k}'} V^d_{\mathbf{k}'-\mathbf{k}} \langle c^\dagger_{b \sigma, \mathbf{k}' \pm \frac{\mathbf{Q}}{2}} 
c_{t \sigma,\mathbf{k}' \pm \frac{\mathbf{Q}}{2}} \rangle, \\
\Delta_\mathbf{k}(\mathbf{Q})
&= \frac{1}{A}\sum\limits_{\mathbf{k}'} V^d_{\mathbf{k}'-\mathbf{k}} \langle c^\dagger_{b \uparrow, \mathbf{k}'-\frac{\mathbf{Q}}{2}} 
c_{t \downarrow,\mathbf{k}'+\frac{\mathbf{Q}}{2}} \rangle, \\
\Delta_\mathbf{k}(-\mathbf{Q})
&= \frac{1}{A}\sum\limits_{\mathbf{k}'} V^d_{\mathbf{k}'-\mathbf{k}} \langle c^\dagger_{b \downarrow, \mathbf{k}'+\frac{\mathbf{Q}}{2}} 
c_{t \uparrow,\mathbf{k}'-\frac{\mathbf{Q}}{2}} \rangle.
\end{split}
\end{equation}
Expressed in the eigenfunctions of Hamiltonian Eq.~(\ref{eq_hamil_4band}),
\begin{equation}
|n \mathbf{k} \rangle = \sum\limits_{l,\sigma} z^{(n)}_{l\sigma}(\mathbf{k}) |l\sigma,\mathbf{k} \rangle,
\end{equation}
the Hamiltonian matrix elements are
\begin{equation}
\label{Eq_mat_ele}
\begin{split}
\varepsilon_{l\sigma,\mathbf{k} \pm \frac{\mathbf{Q}}{2}} &= \frac{\hbar^2}{2m^*} |\mathbf{k}\pm \frac{\mathbf{Q}}{2}|^2, \\
& \quad - \frac{2\pi e^2}{\epsilon_b A}\sum\limits_{n,\mathbf{k}'} \frac{1}{|\mathbf{k}'-\mathbf{k}|} |z^{(n)}_{l\sigma}(\mathbf{k}')|^2 f_{n\mathbf{k}'}, \\
g_{l,\mathbf{k}}(\mathbf{Q}) &= \frac{2\pi e^2}{\epsilon_b A} \sum\limits_{n, \mathbf{k}'}\frac{1}{|\mathbf{k}'-\mathbf{k}|} \bar{z}^{(n)}_{l\uparrow}(\mathbf{k}') z^{(n)}_{l\downarrow}(\mathbf{k}') f_{n\mathbf{k}'}, \\
\Delta_{\mathbf{k}}(0) &= \frac{2\pi e^2}{\epsilon_b A} \sum\limits_{n,\mathbf{k}'} \frac{e^{-|\mathbf{k}'-\mathbf{k}| d}}{|\mathbf{k}'-\mathbf{k}|} \bar{z}^{(n)}_{b \sigma}(\mathbf{k}') z^{(n)}_{t \sigma}(\mathbf{k}') f_{n\mathbf{k}'}, \\
\Delta_{\mathbf{k}}(\mathbf{Q}) &= \frac{2\pi e^2}{\epsilon_b A} \sum\limits_{n,\mathbf{k}'} \frac{e^{-|\mathbf{k}'-\mathbf{k}| d}}{|\mathbf{k}'-\mathbf{k}|} \bar{z}^{(n)}_{b \uparrow}(\mathbf{k}') z^{(n)}_{t \downarrow}(\mathbf{k}') f_{n\mathbf{k}'}, \\
\Delta_{\mathbf{k}}(-\mathbf{Q}) &= \frac{2\pi e^2}{\epsilon_b A} \sum\limits_{n,\mathbf{k}'} \frac{e^{-|\mathbf{k}'-\mathbf{k}| d}}{|\mathbf{k}'-\mathbf{k}|} \bar{z}^{(n)}_{b \downarrow}(\mathbf{k}') z^{(n)}_{t \uparrow}(\mathbf{k}') f_{n\mathbf{k}'}. \\
\end{split}
\end{equation}

Even though we have specified spins associated with the interlayer coherence, the Hamiltonian Eq.~(\ref{eq_hamil_4band}) has SU(2) $\times$ SU(2) symmetry, i.e., independent spin rotational symmetry in each pseudospin sector, and therefore the interlayer order parameters $\Delta_{\mathbf{k}}$ are actually spin independent because of spin-independent Coulomb interaction.

In this four-band model, we have seen the signatures of lowing the total energy by forming SDW and PSDW together near the $S_0$-$S_1$ phase boundary in Ref.~\cite{JZ_interlayerCoherence_2024}. 
However, the density wave momenta for intralayer SDW and interlayer PSDW can be different.

\section{RPA static screening theory with Hubbard-type local field corrections}
\label{Appendix_RPA}
With local field factors $G_{\sigma \sigma'}$, the screened Coulomb potential is
\begin{equation}
\label{Eq_Vsc_appendix}
\begin{split}
V_{\rm sc,\sigma} &= V_{\rm ext, \sigma} + \sum\limits_{\sigma'}v_q(1-G_{\sigma \sigma'}) n_{1\sigma'},
\end{split}
\end{equation}
where $v_q$ is the bare Coulomb potential including the dielectric environment of surrounding media and $n_{1\sigma}$ is the induced charge density,
\begin{equation}
\label{Eq_n1chi0}
\begin{split}
n_{1\sigma} &= \chi_{0\sigma} V_{\rm sc,\sigma} \\
&= \chi_{0\sigma} \big[ V_{\rm ext, \sigma} + \sum\limits_{\sigma'}v_q(1-G_{\sigma \sigma'}) n_{1\sigma'} \big], \\
\end{split}
\end{equation}
which is the response to $V_{\rm sc,\sigma}$ by the non-interacting response function $\chi_{0\sigma}$. 
Absorbing the local field factors $G_{\sigma \sigma'}$ into the response function $\tilde{\chi}_{\sigma}$, $n_{1\sigma}$ can also be written as
\begin{equation}
\label{Eq_n1chitilde}
\begin{split}
n_{1\sigma} &= \tilde{\chi}_{\sigma} \big[ V_{\rm ext, \sigma} + \sum\limits_{\sigma'}v_q n_{1\sigma'} \big].
\end{split}
\end{equation}
Solving the coupled equations (\ref{Eq_n1chi0}) and (\ref{Eq_n1chitilde}), and ignoring the correlation-hole correction, i.e., $G_{\uparrow \downarrow} = G_{\downarrow \uparrow} = 0$,
\begin{equation}
\begin{split}
\tilde{\chi}_{\sigma} = 
\frac{\chi_{0\sigma}}{1 + v_q G_{\sigma\sigma} \chi_{0\sigma}}.
\end{split}
\end{equation}

By solving the coupled equations (\ref{Eq_Vsc_appendix}) and (\ref{Eq_n1chi0})
\begin{equation}
\begin{split}
& V_{\rm sc,\sigma} = V_{\rm ext,\sigma} + \sum\limits_{\sigma'}v_q(1-G_{\sigma\sigma'}) \chi_{0\sigma'} V_{\rm sc,\sigma'} \\
\end{split}
\end{equation}
and in the matrix form
\begin{equation}
\begin{split}
\pmb{V}_{\rm sc} = \pmb{\epsilon}^{-1} \pmb{V}_{\rm ext}.
\end{split}
\end{equation}
The inverse of dielectric constant matrix is
\begin{widetext}
\begin{equation}
\begin{split}
\pmb{\epsilon}^{-1} = \frac{1}{\det(\pmb{\epsilon})} 
\begin{pmatrix}
1-v_q(1-G_{\downarrow\downarrow})\chi_{0\downarrow} & v_q \chi_{0\downarrow} \\
v_q \chi_{0\uparrow} & 1-v_q(1-G_{\uparrow\uparrow})\chi_{0\uparrow}
\end{pmatrix}.
\end{split}
\end{equation}
\end{widetext}
Using $V_{\rm ext,\uparrow} = V_{\rm ext,\downarrow}$, and defining the spin-resolved dielectric constant $\epsilon_\sigma$,
\begin{equation}
\label{Eq_ep_sigma}
\begin{split}
V_{\rm sc,\sigma} &= \frac{V_{\rm ext,\sigma}}{\epsilon_\sigma},
\end{split}
\end{equation}
we have
\begin{equation}
\begin{split}
\epsilon_\sigma = (1+v_q G_{\sigma\sigma} \chi_{0\sigma}) \big[ 1-v_q(\tilde{\chi}_{\uparrow} + \tilde{\chi}_{\downarrow}) \big],
\end{split}
\end{equation}
i.e., Eq.~(\ref{Eq_epsilon_ferro}) in the main text.

\section{The spin ferromagnetic phase with RPA static screening}
\label{Appendix_RPA_spin}
With the primitive RPA static screening, the spin paramagnetic state is stable for all $r_s$ values considered in our calculations. It is known that RPA static screening theory always overestimates the strength of screening because it ignores the exchange-correlation effects.
In magnetic transitions, the exchange interaction is important. We, therefore, consider the Hubbard-type local field factor to the primitive RPA static theory to include the corrections from the exchange effect.

\begin{figure}
\centering
\includegraphics[width=1.0\columnwidth]{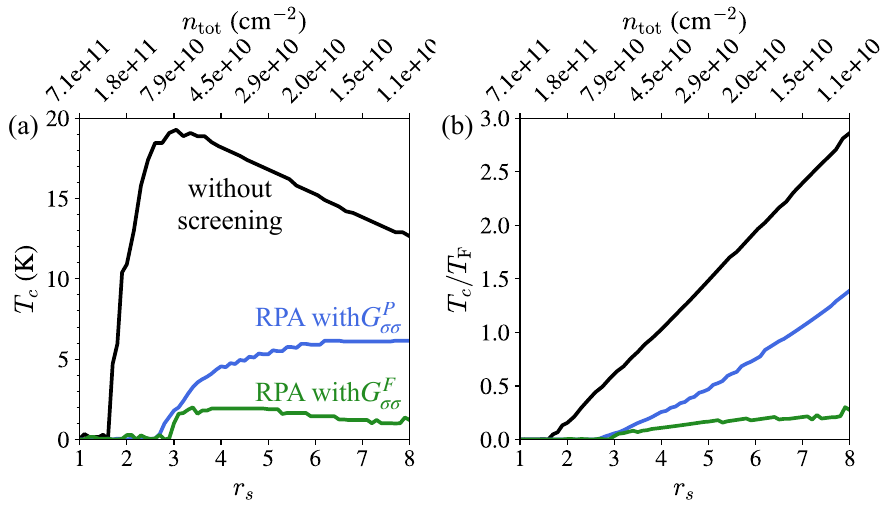}
\caption{\label{fig_S1_RPA} {
(a) $T_c$ of the spin polarized state, calculated by finite-temperature self-consistent HF. (b) Same as (a) but $T_c$ is shown in the unit of Fermi energy $T_{\rm F} = \varepsilon_{\rm F}/k_{\rm B}$. 
Without screenings (black lines), $T_c$ is the highest and the critical $r_{s,c}$ of the paramagnetic to ferromagnetic transition is $\sim 2$ (the Bloch transition). 
With the RPA static screening including the Hubbard-type paramagnetic local field correction ($G_{\sigma \sigma}^P$, blue lines), $T_c$ is suppressed and the critical $r_{s,c}$ is pushed to a higher value $\lesssim 3$.
$T_c$ is further suppressed and $r_{s,c} \sim 3$ including the Hubbard-type ferromagnetic local field correction ($G_{\sigma \sigma}^F$, green lines), which takes into account different Fermi surfaces of two spins. 
Under the primitive RPA static screening theory, spin is not ordered for all $r_s$ values considered in our calculations.  
  }}
\end{figure}

Figure~\ref{fig_S1_RPA} shows $T_c$ of the spin polarized state, calculated by finite-temperature self-consistent HF.
Without screenings (black lines), $T_c$ is the highest and the critical $r_{s,c}$ of the paramagnetic to ferromagnetic transition is approximately $\sim 2$ (the Bloch transition). 
With the RPA static screening including the Hubbard-type paramagnetic local field correction ($G_{\sigma \sigma}^P$, blue lines), $T_c$ is suppressed and the critical $r_{s,c}$ is pushed to a higher value $\lesssim 3$.
$T_c$ is further suppressed and $r_{s,c} \sim 3$ including the Hubbard-type ferromagnetic local field correction ($G_{\sigma \sigma}^F$, green lines), which takes into account different Fermi surfaces of two spins. 
Under the primitive RPA static screening theory, spin is not ordered for all $r_s$ values considered in our calculations.


\begin{thebibliography}{65}%
\makeatletter
\providecommand \@ifxundefined [1]{%
 \@ifx{#1\undefined}
}%
\providecommand \@ifnum [1]{%
 \ifnum #1\expandafter \@firstoftwo
 \else \expandafter \@secondoftwo
 \fi
}%
\providecommand \@ifx [1]{%
 \ifx #1\expandafter \@firstoftwo
 \else \expandafter \@secondoftwo
 \fi
}%
\providecommand \natexlab [1]{#1}%
\providecommand \enquote  [1]{``#1''}%
\providecommand \bibnamefont  [1]{#1}%
\providecommand \bibfnamefont [1]{#1}%
\providecommand \citenamefont [1]{#1}%
\providecommand \href@noop [0]{\@secondoftwo}%
\providecommand \href [0]{\begingroup \@sanitize@url \@href}%
\providecommand \@href[1]{\@@startlink{#1}\@@href}%
\providecommand \@@href[1]{\endgroup#1\@@endlink}%
\providecommand \@sanitize@url [0]{\catcode `\\12\catcode `\$12\catcode `\&12\catcode `\#12\catcode `\^12\catcode `\_12\catcode `\%12\relax}%
\providecommand \@@startlink[1]{}%
\providecommand \@@endlink[0]{}%
\providecommand \url  [0]{\begingroup\@sanitize@url \@url }%
\providecommand \@url [1]{\endgroup\@href {#1}{\urlprefix }}%
\providecommand \urlprefix  [0]{URL }%
\providecommand \Eprint [0]{\href }%
\providecommand \doibase [0]{http://dx.doi.org/}%
\providecommand \selectlanguage [0]{\@gobble}%
\providecommand \bibinfo  [0]{\@secondoftwo}%
\providecommand \bibfield  [0]{\@secondoftwo}%
\providecommand \translation [1]{[#1]}%
\providecommand \BibitemOpen [0]{}%
\providecommand \bibitemStop [0]{}%
\providecommand \bibitemNoStop [0]{.\EOS\space}%
\providecommand \EOS [0]{\spacefactor3000\relax}%
\providecommand \BibitemShut  [1]{\csname bibitem#1\endcsname}%
\let\auto@bib@innerbib\@empty
\bibitem [{\citenamefont {Zheng}\ \emph {et~al.}(1997)\citenamefont {Zheng}, \citenamefont {Ortalano},\ and\ \citenamefont {Das~Sarma}}]{LZheng_doubleQW_1997}%
  \BibitemOpen
  \bibfield  {author} {\bibinfo {author} {\bibfnamefont {L.}~\bibnamefont {Zheng}}, \bibinfo {author} {\bibfnamefont {M.~W.}\ \bibnamefont {Ortalano}}, \ and\ \bibinfo {author} {\bibfnamefont {S.}~\bibnamefont {Das~Sarma}},\ }\bibfield  {title} {\emph {\bibinfo {title} {Exchange instabilities in semiconductor double-quantum-well systems},\ }}\href {\doibase 10.1103/PhysRevB.55.4506} {\bibfield  {journal} {\bibinfo  {journal} {Phys. Rev. B}\ }\textbf {\bibinfo {volume} {55}},\ \bibinfo {pages} {4506} (\bibinfo {year} {1997})}\BibitemShut {NoStop}%
\bibitem [{\citenamefont {Das~Sarma}\ \emph {et~al.}(1998)\citenamefont {Das~Sarma}, \citenamefont {Ortalano},\ and\ \citenamefont {Zheng}}]{DasSarma_doubleQW_1998}%
  \BibitemOpen
  \bibfield  {author} {\bibinfo {author} {\bibfnamefont {S.}~\bibnamefont {Das~Sarma}}, \bibinfo {author} {\bibfnamefont {M.~W.}\ \bibnamefont {Ortalano}}, \ and\ \bibinfo {author} {\bibfnamefont {L.}~\bibnamefont {Zheng}},\ }\bibfield  {title} {\emph {\bibinfo {title} {Bilayer to monolayer charge-transfer instability in semiconductor double-quantum-well structures},\ }}\href {\doibase 10.1103/PhysRevB.58.7453} {\bibfield  {journal} {\bibinfo  {journal} {Phys. Rev. B}\ }\textbf {\bibinfo {volume} {58}},\ \bibinfo {pages} {7453} (\bibinfo {year} {1998})}\BibitemShut {NoStop}%
\bibitem [{\citenamefont {Zhu}\ and\ \citenamefont {Das~Sarma}(2024)}]{JZ_interlayerCoherence_2024}%
  \BibitemOpen
  \bibfield  {author} {\bibinfo {author} {\bibfnamefont {J.}~\bibnamefont {Zhu}}\ and\ \bibinfo {author} {\bibfnamefont {S.}~\bibnamefont {Das~Sarma}},\ }\bibfield  {title} {\emph {\bibinfo {title} {Interaction and coherence in two-dimensional bilayers},\ }}\href {\doibase 10.1103/PhysRevB.109.085129} {\bibfield  {journal} {\bibinfo  {journal} {Phys. Rev. B}\ }\textbf {\bibinfo {volume} {109}},\ \bibinfo {pages} {085129} (\bibinfo {year} {2024})}\BibitemShut {NoStop}%
\bibitem [{\citenamefont {Cookmeyer}\ and\ \citenamefont {Das~Sarma}(2024)}]{Tessa_2Dbilayer_2024}%
  \BibitemOpen
  \bibfield  {author} {\bibinfo {author} {\bibfnamefont {T.}~\bibnamefont {Cookmeyer}}\ and\ \bibinfo {author} {\bibfnamefont {S.}~\bibnamefont {Das~Sarma}},\ }\bibfield  {title} {\emph {\bibinfo {title} {Symmetry breaking in zero-field two-dimensional electron bilayers},\ }}\href {\doibase 10.1103/PhysRevB.109.115307} {\bibfield  {journal} {\bibinfo  {journal} {Phys. Rev. B}\ }\textbf {\bibinfo {volume} {109}},\ \bibinfo {pages} {115307} (\bibinfo {year} {2024})}\BibitemShut {NoStop}%
\bibitem [{\citenamefont {Stern}\ \emph {et~al.}(2000)\citenamefont {Stern}, \citenamefont {Das~Sarma}, \citenamefont {Fisher},\ and\ \citenamefont {Girvin}}]{XYbilayer_Girvin_2000}%
  \BibitemOpen
  \bibfield  {author} {\bibinfo {author} {\bibfnamefont {A.}~\bibnamefont {Stern}}, \bibinfo {author} {\bibfnamefont {S.}~\bibnamefont {Das~Sarma}}, \bibinfo {author} {\bibfnamefont {M.~P.~A.}\ \bibnamefont {Fisher}}, \ and\ \bibinfo {author} {\bibfnamefont {S.~M.}\ \bibnamefont {Girvin}},\ }\bibfield  {title} {\emph {\bibinfo {title} {Dissipationless transport in low-density bilayer systems},\ }}\href {\doibase 10.1103/PhysRevLett.84.139} {\bibfield  {journal} {\bibinfo  {journal} {Phys. Rev. Lett.}\ }\textbf {\bibinfo {volume} {84}},\ \bibinfo {pages} {139} (\bibinfo {year} {2000})}\BibitemShut {NoStop}%
\bibitem [{\citenamefont {Radtke}\ and\ \citenamefont {{Das Sarma}}(1996)}]{SDS_2DbilayerSpin_1996}%
  \BibitemOpen
  \bibfield  {author} {\bibinfo {author} {\bibfnamefont {R.}~\bibnamefont {Radtke}}\ and\ \bibinfo {author} {\bibfnamefont {S.}~\bibnamefont {{Das Sarma}}},\ }\bibfield  {title} {\emph {\bibinfo {title} {Collective modes in a symmetry-broken phase: Antiferromagnetically correlated quantum wells},\ }}\href {\doibase https://doi.org/10.1016/0038-1098(96)00200-1} {\bibfield  {journal} {\bibinfo  {journal} {Solid State Communications}\ }\textbf {\bibinfo {volume} {98}},\ \bibinfo {pages} {771} (\bibinfo {year} {1996})}\BibitemShut {NoStop}%
\bibitem [{\citenamefont {Radtke}\ \emph {et~al.}(1998)\citenamefont {Radtke}, \citenamefont {Das~Sarma},\ and\ \citenamefont {MacDonald}}]{AHM_2DbilayerSpin_1998}%
  \BibitemOpen
  \bibfield  {author} {\bibinfo {author} {\bibfnamefont {R.~J.}\ \bibnamefont {Radtke}}, \bibinfo {author} {\bibfnamefont {S.}~\bibnamefont {Das~Sarma}}, \ and\ \bibinfo {author} {\bibfnamefont {A.~H.}\ \bibnamefont {MacDonald}},\ }\bibfield  {title} {\emph {\bibinfo {title} {Mode mixing in antiferromagnetically correlated double quantum wells},\ }}\href {\doibase 10.1103/PhysRevB.57.2342} {\bibfield  {journal} {\bibinfo  {journal} {Phys. Rev. B}\ }\textbf {\bibinfo {volume} {57}},\ \bibinfo {pages} {2342} (\bibinfo {year} {1998})}\BibitemShut {NoStop}%
\bibitem [{\citenamefont {Radtke}\ \emph {et~al.}(1996)\citenamefont {Radtke}, \citenamefont {Tamborenea},\ and\ \citenamefont {Das~Sarma}}]{Radtke_2DbilayerSpin_1996}%
  \BibitemOpen
  \bibfield  {author} {\bibinfo {author} {\bibfnamefont {R.~J.}\ \bibnamefont {Radtke}}, \bibinfo {author} {\bibfnamefont {P.~I.}\ \bibnamefont {Tamborenea}}, \ and\ \bibinfo {author} {\bibfnamefont {S.}~\bibnamefont {Das~Sarma}},\ }\bibfield  {title} {\emph {\bibinfo {title} {Spin instabilities in coupled semiconductor quantum wells},\ }}\href {\doibase 10.1103/PhysRevB.54.13832} {\bibfield  {journal} {\bibinfo  {journal} {Phys. Rev. B}\ }\textbf {\bibinfo {volume} {54}},\ \bibinfo {pages} {13832} (\bibinfo {year} {1996})}\BibitemShut {NoStop}%
\bibitem [{\citenamefont {Overhauser}(1960)}]{Overhauser_SDW_1960}%
  \BibitemOpen
  \bibfield  {author} {\bibinfo {author} {\bibfnamefont {A.~W.}\ \bibnamefont {Overhauser}},\ }\bibfield  {title} {\emph {\bibinfo {title} {Giant spin density waves},\ }}\href {\doibase 10.1103/PhysRevLett.4.462} {\bibfield  {journal} {\bibinfo  {journal} {Phys. Rev. Lett.}\ }\textbf {\bibinfo {volume} {4}},\ \bibinfo {pages} {462} (\bibinfo {year} {1960})}\BibitemShut {NoStop}%
\bibitem [{\citenamefont {Overhauser}(1962)}]{Overhauser_SDW_1962}%
  \BibitemOpen
  \bibfield  {author} {\bibinfo {author} {\bibfnamefont {A.~W.}\ \bibnamefont {Overhauser}},\ }\bibfield  {title} {\emph {\bibinfo {title} {Spin density waves in an electron gas},\ }}\href {\doibase 10.1103/PhysRev.128.1437} {\bibfield  {journal} {\bibinfo  {journal} {Phys. Rev.}\ }\textbf {\bibinfo {volume} {128}},\ \bibinfo {pages} {1437} (\bibinfo {year} {1962})}\BibitemShut {NoStop}%
\bibitem [{\citenamefont {Kohn}\ and\ \citenamefont {Nettel}(1960)}]{Kohn_SDW_1960}%
  \BibitemOpen
  \bibfield  {author} {\bibinfo {author} {\bibfnamefont {W.}~\bibnamefont {Kohn}}\ and\ \bibinfo {author} {\bibfnamefont {S.~J.}\ \bibnamefont {Nettel}},\ }\bibfield  {title} {\emph {\bibinfo {title} {Giant fluctuations in a degenerate {Fermi} gas},\ }}\href {\doibase 10.1103/PhysRevLett.5.8} {\bibfield  {journal} {\bibinfo  {journal} {Phys. Rev. Lett.}\ }\textbf {\bibinfo {volume} {5}},\ \bibinfo {pages} {8} (\bibinfo {year} {1960})}\BibitemShut {NoStop}%
\bibitem [{\citenamefont {Fedders}\ and\ \citenamefont {Martin}(1966)}]{Fedders_correlation_1966}%
  \BibitemOpen
  \bibfield  {author} {\bibinfo {author} {\bibfnamefont {P.~A.}\ \bibnamefont {Fedders}}\ and\ \bibinfo {author} {\bibfnamefont {P.~C.}\ \bibnamefont {Martin}},\ }\bibfield  {title} {\emph {\bibinfo {title} {Itinerant antiferromagnetism},\ }}\href {\doibase 10.1103/PhysRev.143.245} {\bibfield  {journal} {\bibinfo  {journal} {Phys. Rev.}\ }\textbf {\bibinfo {volume} {143}},\ \bibinfo {pages} {245} (\bibinfo {year} {1966})}\BibitemShut {NoStop}%
\bibitem [{\citenamefont {Ando}(1974)}]{Ando_QOscill_1974}%
  \BibitemOpen
  \bibfield  {author} {\bibinfo {author} {\bibfnamefont {T.}~\bibnamefont {Ando}},\ }\bibfield  {title} {\emph {\bibinfo {title} {Theory of quantum transport in a two-dimensional electron system under magnetic fields. {IV}. oscillatory conductivity},\ }}\href {\doibase 10.1143/JPSJ.37.1233} {\bibfield  {journal} {\bibinfo  {journal} {Journal of the Physical Society of Japan}\ }\textbf {\bibinfo {volume} {37}},\ \bibinfo {pages} {1233} (\bibinfo {year} {1974})}\BibitemShut {NoStop}%
\bibitem [{\citenamefont {Smr\ifmmode~\check{c}\else \v{c}\fi{}ka}\ \emph {et~al.}(1995)\citenamefont {Smr\ifmmode~\check{c}\else \v{c}\fi{}ka}, \citenamefont {Va\ifmmode~\check{s}\else \v{s}\fi{}ek}, \citenamefont {Kol\'a\ifmmode~\check{c}\else \v{c}\fi{}ek}, \citenamefont {Jungwirth},\ and\ \citenamefont {Cukr}}]{Smrcka_CyclotronMass_1995}%
  \BibitemOpen
  \bibfield  {author} {\bibinfo {author} {\bibfnamefont {L.}~\bibnamefont {Smr\ifmmode~\check{c}\else \v{c}\fi{}ka}}, \bibinfo {author} {\bibfnamefont {P.}~\bibnamefont {Va\ifmmode~\check{s}\else \v{s}\fi{}ek}}, \bibinfo {author} {\bibfnamefont {J.}~\bibnamefont {Kol\'a\ifmmode~\check{c}\else \v{c}\fi{}ek}}, \bibinfo {author} {\bibfnamefont {T.}~\bibnamefont {Jungwirth}}, \ and\ \bibinfo {author} {\bibfnamefont {M.}~\bibnamefont {Cukr}},\ }\bibfield  {title} {\emph {\bibinfo {title} {Cyclotron effective mass of a two-dimensional electron layer at the {GaAs/${\mathrm{Al}}_{\mathit{x}}$${\mathrm{Ga}}_{1\mathrm{\ensuremath{-}}\mathit{x}}$As} heterojunction subject to in-plane magnetic fields},\ }}\href {\doibase 10.1103/PhysRevB.51.18011} {\bibfield  {journal} {\bibinfo  {journal} {Phys. Rev. B}\ }\textbf {\bibinfo {volume} {51}},\ \bibinfo {pages} {18011} (\bibinfo {year} {1995})}\BibitemShut {NoStop}%
\bibitem [{\citenamefont {Kaganov}\ and\ \citenamefont {Lifshits}(1979)}]{Kaganov_Lifshits_1979}%
  \BibitemOpen
  \bibfield  {author} {\bibinfo {author} {\bibfnamefont {M.~I.}\ \bibnamefont {Kaganov}}\ and\ \bibinfo {author} {\bibfnamefont {I.~M.}\ \bibnamefont {Lifshits}},\ }\bibfield  {title} {\emph {\bibinfo {title} {Electron theory of metals and geometry},\ }}\href {\doibase 10.1070/PU1979v022n11ABEH005648} {\bibfield  {journal} {\bibinfo  {journal} {Soviet Physics Uspekhi}\ }\textbf {\bibinfo {volume} {22}},\ \bibinfo {pages} {904} (\bibinfo {year} {1979})}\BibitemShut {NoStop}%
\bibitem [{\citenamefont {Cole}\ \emph {et~al.}(1977)\citenamefont {Cole}, \citenamefont {Lakhani},\ and\ \citenamefont {Stiles}}]{Cole_tiltedSi_1977}%
  \BibitemOpen
  \bibfield  {author} {\bibinfo {author} {\bibfnamefont {T.}~\bibnamefont {Cole}}, \bibinfo {author} {\bibfnamefont {A.~A.}\ \bibnamefont {Lakhani}}, \ and\ \bibinfo {author} {\bibfnamefont {P.~J.}\ \bibnamefont {Stiles}},\ }\bibfield  {title} {\emph {\bibinfo {title} {Influence of a one-dimensional superlattice on a two-dimensional electron gas},\ }}\href {\doibase 10.1103/PhysRevLett.38.722} {\bibfield  {journal} {\bibinfo  {journal} {Phys. Rev. Lett.}\ }\textbf {\bibinfo {volume} {38}},\ \bibinfo {pages} {722} (\bibinfo {year} {1977})}\BibitemShut {NoStop}%
\bibitem [{\citenamefont {Tsui}\ \emph {et~al.}(1978)\citenamefont {Tsui}, \citenamefont {Sturge}, \citenamefont {Kamgar},\ and\ \citenamefont {Allen}}]{Tsui_tiltedSi_1978}%
  \BibitemOpen
  \bibfield  {author} {\bibinfo {author} {\bibfnamefont {D.~C.}\ \bibnamefont {Tsui}}, \bibinfo {author} {\bibfnamefont {M.~D.}\ \bibnamefont {Sturge}}, \bibinfo {author} {\bibfnamefont {A.}~\bibnamefont {Kamgar}}, \ and\ \bibinfo {author} {\bibfnamefont {S.~J.}\ \bibnamefont {Allen}},\ }\bibfield  {title} {\emph {\bibinfo {title} {Surface band structure of electron inversion layers on vicinal planes of {Si}(100)},\ }}\href {\doibase 10.1103/PhysRevLett.40.1667} {\bibfield  {journal} {\bibinfo  {journal} {Phys. Rev. Lett.}\ }\textbf {\bibinfo {volume} {40}},\ \bibinfo {pages} {1667} (\bibinfo {year} {1978})}\BibitemShut {NoStop}%
\bibitem [{\citenamefont {Sesselmann}\ and\ \citenamefont {Kotthaus}(1979)}]{Sesselmann_OpticalGap_1979}%
  \BibitemOpen
  \bibfield  {author} {\bibinfo {author} {\bibfnamefont {W.}~\bibnamefont {Sesselmann}}\ and\ \bibinfo {author} {\bibfnamefont {J.}~\bibnamefont {Kotthaus}},\ }\bibfield  {title} {\emph {\bibinfo {title} {Spectroscopic determination of the energy gaps in the inversion layer band structure on vicinal planes of {Si}(001)},\ }}\href {\doibase https://doi.org/10.1016/0038-1098(79)90433-2} {\bibfield  {journal} {\bibinfo  {journal} {Solid State Communications}\ }\textbf {\bibinfo {volume} {31}},\ \bibinfo {pages} {193} (\bibinfo {year} {1979})}\BibitemShut {NoStop}%
\bibitem [{\citenamefont {Kamgar}\ \emph {et~al.}(1980)\citenamefont {Kamgar}, \citenamefont {Sturge},\ and\ \citenamefont {Tsui}}]{Kamgar_OpticalGap_1980}%
  \BibitemOpen
  \bibfield  {author} {\bibinfo {author} {\bibfnamefont {A.}~\bibnamefont {Kamgar}}, \bibinfo {author} {\bibfnamefont {M.~D.}\ \bibnamefont {Sturge}}, \ and\ \bibinfo {author} {\bibfnamefont {D.~C.}\ \bibnamefont {Tsui}},\ }\bibfield  {title} {\emph {\bibinfo {title} {Optical measurements of the minigaps in electron inversion layers on vicinal planes of {Si}(001)},\ }}\href {\doibase 10.1103/PhysRevB.22.841} {\bibfield  {journal} {\bibinfo  {journal} {Phys. Rev. B}\ }\textbf {\bibinfo {volume} {22}},\ \bibinfo {pages} {841} (\bibinfo {year} {1980})}\BibitemShut {NoStop}%
\bibitem [{\citenamefont {Matheson}\ and\ \citenamefont {Higgins}(1982)}]{Matheson_tiltedSi_1982}%
  \BibitemOpen
  \bibfield  {author} {\bibinfo {author} {\bibfnamefont {T.~G.}\ \bibnamefont {Matheson}}\ and\ \bibinfo {author} {\bibfnamefont {R.~J.}\ \bibnamefont {Higgins}},\ }\bibfield  {title} {\emph {\bibinfo {title} {Tunneling in tilted {Si} inversion layers},\ }}\href {\doibase 10.1103/PhysRevB.25.2633} {\bibfield  {journal} {\bibinfo  {journal} {Phys. Rev. B}\ }\textbf {\bibinfo {volume} {25}},\ \bibinfo {pages} {2633} (\bibinfo {year} {1982})}\BibitemShut {NoStop}%
\bibitem [{\citenamefont {Simmons}\ \emph {et~al.}(1994)\citenamefont {Simmons}, \citenamefont {Lyo}, \citenamefont {Harff},\ and\ \citenamefont {Klem}}]{Simmons_dQWinBpara_1994}%
  \BibitemOpen
  \bibfield  {author} {\bibinfo {author} {\bibfnamefont {J.~A.}\ \bibnamefont {Simmons}}, \bibinfo {author} {\bibfnamefont {S.~K.}\ \bibnamefont {Lyo}}, \bibinfo {author} {\bibfnamefont {N.~E.}\ \bibnamefont {Harff}}, \ and\ \bibinfo {author} {\bibfnamefont {J.~F.}\ \bibnamefont {Klem}},\ }\bibfield  {title} {\emph {\bibinfo {title} {Conductance modulation in double quantum wells due to magnetic field-induced anticrossing},\ }}\href {\doibase 10.1103/PhysRevLett.73.2256} {\bibfield  {journal} {\bibinfo  {journal} {Phys. Rev. Lett.}\ }\textbf {\bibinfo {volume} {73}},\ \bibinfo {pages} {2256} (\bibinfo {year} {1994})}\BibitemShut {NoStop}%
\bibitem [{\citenamefont {Simmons}\ \emph {et~al.}(1995)\citenamefont {Simmons}, \citenamefont {Harff},\ and\ \citenamefont {Klem}}]{Simmons_dQWinBpara_1995}%
  \BibitemOpen
  \bibfield  {author} {\bibinfo {author} {\bibfnamefont {J.~A.}\ \bibnamefont {Simmons}}, \bibinfo {author} {\bibfnamefont {N.~E.}\ \bibnamefont {Harff}}, \ and\ \bibinfo {author} {\bibfnamefont {J.~F.}\ \bibnamefont {Klem}},\ }\bibfield  {title} {\emph {\bibinfo {title} {Observation of extreme field-induced mass deviations in double quantum wells},\ }}\href {\doibase 10.1103/PhysRevB.51.11156} {\bibfield  {journal} {\bibinfo  {journal} {Phys. Rev. B}\ }\textbf {\bibinfo {volume} {51}},\ \bibinfo {pages} {11156} (\bibinfo {year} {1995})}\BibitemShut {NoStop}%
\bibitem [{\citenamefont {Kurobe}\ \emph {et~al.}(1994)\citenamefont {Kurobe}, \citenamefont {Castleton}, \citenamefont {Linfield}, \citenamefont {Grimshaw}, \citenamefont {Brown}, \citenamefont {Ritchie}, \citenamefont {Pepper},\ and\ \citenamefont {Jones}}]{Kurobe_dQWinBpara_1994}%
  \BibitemOpen
  \bibfield  {author} {\bibinfo {author} {\bibfnamefont {A.}~\bibnamefont {Kurobe}}, \bibinfo {author} {\bibfnamefont {I.~M.}\ \bibnamefont {Castleton}}, \bibinfo {author} {\bibfnamefont {E.~H.}\ \bibnamefont {Linfield}}, \bibinfo {author} {\bibfnamefont {M.~P.}\ \bibnamefont {Grimshaw}}, \bibinfo {author} {\bibfnamefont {K.~M.}\ \bibnamefont {Brown}}, \bibinfo {author} {\bibfnamefont {D.~A.}\ \bibnamefont {Ritchie}}, \bibinfo {author} {\bibfnamefont {M.}~\bibnamefont {Pepper}}, \ and\ \bibinfo {author} {\bibfnamefont {G.~A.~C.}\ \bibnamefont {Jones}},\ }\bibfield  {title} {\emph {\bibinfo {title} {Wave functions and {Fermi} surfaces of strongly coupled two-dimensional electron gases investigated by in-plane magnetoresistance},\ }}\href {\doibase 10.1103/PhysRevB.50.4889} {\bibfield  {journal} {\bibinfo  {journal} {Phys. Rev. B}\ }\textbf {\bibinfo {volume} {50}},\ \bibinfo {pages} {4889} (\bibinfo {year} {1994})}\BibitemShut {NoStop}%
\bibitem [{\citenamefont {Lyo}(1994)}]{Lyo_dQWinBpara_1994}%
  \BibitemOpen
  \bibfield  {author} {\bibinfo {author} {\bibfnamefont {S.~K.}\ \bibnamefont {Lyo}},\ }\bibfield  {title} {\emph {\bibinfo {title} {Transport and level anticrossing in strongly coupled double quantum wells with in-plane magnetic fields},\ }}\href {\doibase 10.1103/PhysRevB.50.4965} {\bibfield  {journal} {\bibinfo  {journal} {Phys. Rev. B}\ }\textbf {\bibinfo {volume} {50}},\ \bibinfo {pages} {4965} (\bibinfo {year} {1994})}\BibitemShut {NoStop}%
\bibitem [{\citenamefont {Lyo}(1995)}]{Lyo_dQWinBpara_1995}%
  \BibitemOpen
  \bibfield  {author} {\bibinfo {author} {\bibfnamefont {S.~K.}\ \bibnamefont {Lyo}},\ }\bibfield  {title} {\emph {\bibinfo {title} {Giant field-induced variation of the cyclotron mass in coupled two-dimensional electron gases},\ }}\href {\doibase 10.1103/PhysRevB.51.11160} {\bibfield  {journal} {\bibinfo  {journal} {Phys. Rev. B}\ }\textbf {\bibinfo {volume} {51}},\ \bibinfo {pages} {11160} (\bibinfo {year} {1995})}\BibitemShut {NoStop}%
\bibitem [{\citenamefont {Ando}\ \emph {et~al.}(1982)\citenamefont {Ando}, \citenamefont {Fowler},\ and\ \citenamefont {Stern}}]{Ando_minigap_review_1982}%
  \BibitemOpen
  \bibfield  {author} {\bibinfo {author} {\bibfnamefont {T.}~\bibnamefont {Ando}}, \bibinfo {author} {\bibfnamefont {A.~B.}\ \bibnamefont {Fowler}}, \ and\ \bibinfo {author} {\bibfnamefont {F.}~\bibnamefont {Stern}},\ }\bibfield  {title} {\emph {\bibinfo {title} {Electronic properties of two-dimensional systems},\ }}\href {\doibase 10.1103/RevModPhys.54.437} {\bibfield  {journal} {\bibinfo  {journal} {Rev. Mod. Phys.}\ }\textbf {\bibinfo {volume} {54}},\ \bibinfo {pages} {437} (\bibinfo {year} {1982})}\BibitemShut {NoStop}%
\bibitem [{\citenamefont {Sham}\ \emph {et~al.}(1978)\citenamefont {Sham}, \citenamefont {Allen}, \citenamefont {Kamgar},\ and\ \citenamefont {Tsui}}]{Sham_tiltedSi_1978}%
  \BibitemOpen
  \bibfield  {author} {\bibinfo {author} {\bibfnamefont {L.~J.}\ \bibnamefont {Sham}}, \bibinfo {author} {\bibfnamefont {S.~J.}\ \bibnamefont {Allen}}, \bibinfo {author} {\bibfnamefont {A.}~\bibnamefont {Kamgar}}, \ and\ \bibinfo {author} {\bibfnamefont {D.~C.}\ \bibnamefont {Tsui}},\ }\bibfield  {title} {\emph {\bibinfo {title} {Valley-valley splitting in inversion layers on a high-index surface of silicon},\ }}\href {\doibase 10.1103/PhysRevLett.40.472} {\bibfield  {journal} {\bibinfo  {journal} {Phys. Rev. Lett.}\ }\textbf {\bibinfo {volume} {40}},\ \bibinfo {pages} {472} (\bibinfo {year} {1978})}\BibitemShut {NoStop}%
\bibitem [{\citenamefont {Winkler}\ \emph {et~al.}(1989)\citenamefont {Winkler}, \citenamefont {Kotthaus},\ and\ \citenamefont {Ploog}}]{Winkler_1Dsuperlattice_1989}%
  \BibitemOpen
  \bibfield  {author} {\bibinfo {author} {\bibfnamefont {R.~W.}\ \bibnamefont {Winkler}}, \bibinfo {author} {\bibfnamefont {J.~P.}\ \bibnamefont {Kotthaus}}, \ and\ \bibinfo {author} {\bibfnamefont {K.}~\bibnamefont {Ploog}},\ }\bibfield  {title} {\emph {\bibinfo {title} {Landau band conductivity in a two-dimensional electron system modulated by an artificial one-dimensional superlattice potential},\ }}\href {\doibase 10.1103/PhysRevLett.62.1177} {\bibfield  {journal} {\bibinfo  {journal} {Phys. Rev. Lett.}\ }\textbf {\bibinfo {volume} {62}},\ \bibinfo {pages} {1177} (\bibinfo {year} {1989})}\BibitemShut {NoStop}%
\bibitem [{\citenamefont {Sze}(1981)}]{shklovskii2013}%
  \BibitemOpen
  \bibfield  {author} {\bibinfo {author} {\bibfnamefont {S.~M.}\ \bibnamefont {Sze}},\ }\href@noop {} {\emph {\bibinfo {title} {Physics of Semiconductor Devices}}}\ (\bibinfo  {publisher} {Wiley, New York},\ \bibinfo {year} {1981})\ p.\ \bibinfo {pages} {850}\BibitemShut {NoStop}%
\bibitem [{\citenamefont {Zhang}\ and\ \citenamefont {Ceperley}(2008)}]{Zhang_SDW_2008}%
  \BibitemOpen
  \bibfield  {author} {\bibinfo {author} {\bibfnamefont {S.}~\bibnamefont {Zhang}}\ and\ \bibinfo {author} {\bibfnamefont {D.~M.}\ \bibnamefont {Ceperley}},\ }\bibfield  {title} {\emph {\bibinfo {title} {{Hartree-Fock} ground state of the three-dimensional electron gas},\ }}\href {\doibase 10.1103/PhysRevLett.100.236404} {\bibfield  {journal} {\bibinfo  {journal} {Phys. Rev. Lett.}\ }\textbf {\bibinfo {volume} {100}},\ \bibinfo {pages} {236404} (\bibinfo {year} {2008})}\BibitemShut {NoStop}%
\bibitem [{\citenamefont {Giuliani}\ and\ \citenamefont {Vignale}(2008)}]{Giuliani_SDW_2008}%
  \BibitemOpen
  \bibfield  {author} {\bibinfo {author} {\bibfnamefont {G.~F.}\ \bibnamefont {Giuliani}}\ and\ \bibinfo {author} {\bibfnamefont {G.}~\bibnamefont {Vignale}},\ }\bibfield  {title} {\emph {\bibinfo {title} {Absence of certain exchange driven instabilities of an electron gas at high densities},\ }}\href {\doibase 10.1103/PhysRevB.78.075110} {\bibfield  {journal} {\bibinfo  {journal} {Phys. Rev. B}\ }\textbf {\bibinfo {volume} {78}},\ \bibinfo {pages} {075110} (\bibinfo {year} {2008})}\BibitemShut {NoStop}%
\bibitem [{\citenamefont {Kurth}\ and\ \citenamefont {Eich}(2009)}]{Kurth_SDW_SDFT_2009}%
  \BibitemOpen
  \bibfield  {author} {\bibinfo {author} {\bibfnamefont {S.}~\bibnamefont {Kurth}}\ and\ \bibinfo {author} {\bibfnamefont {F.~G.}\ \bibnamefont {Eich}},\ }\bibfield  {title} {\emph {\bibinfo {title} {Overhauser's spin-density wave in exact-exchange spin-density functional theory},\ }}\href {\doibase 10.1103/PhysRevB.80.125120} {\bibfield  {journal} {\bibinfo  {journal} {Phys. Rev. B}\ }\textbf {\bibinfo {volume} {80}},\ \bibinfo {pages} {125120} (\bibinfo {year} {2009})}\BibitemShut {NoStop}%
\bibitem [{\citenamefont {Delyon}\ \emph {et~al.}(2015)\citenamefont {Delyon}, \citenamefont {Bernu}, \citenamefont {Baguet},\ and\ \citenamefont {Holzmann}}]{Delyon_SDW_2015}%
  \BibitemOpen
  \bibfield  {author} {\bibinfo {author} {\bibfnamefont {F.}~\bibnamefont {Delyon}}, \bibinfo {author} {\bibfnamefont {B.}~\bibnamefont {Bernu}}, \bibinfo {author} {\bibfnamefont {L.}~\bibnamefont {Baguet}}, \ and\ \bibinfo {author} {\bibfnamefont {M.}~\bibnamefont {Holzmann}},\ }\bibfield  {title} {\emph {\bibinfo {title} {Upper bounds of spin-density wave energies in the homogeneous electron gas},\ }}\href {\doibase 10.1103/PhysRevB.92.235124} {\bibfield  {journal} {\bibinfo  {journal} {Phys. Rev. B}\ }\textbf {\bibinfo {volume} {92}},\ \bibinfo {pages} {235124} (\bibinfo {year} {2015})}\BibitemShut {NoStop}%
\bibitem [{\citenamefont {Gontier}\ \emph {et~al.}(2019)\citenamefont {Gontier}, \citenamefont {Hainzl},\ and\ \citenamefont {Lewin}}]{Gontier_SDW_2019}%
  \BibitemOpen
  \bibfield  {author} {\bibinfo {author} {\bibfnamefont {D.}~\bibnamefont {Gontier}}, \bibinfo {author} {\bibfnamefont {C.}~\bibnamefont {Hainzl}}, \ and\ \bibinfo {author} {\bibfnamefont {M.}~\bibnamefont {Lewin}},\ }\bibfield  {title} {\emph {\bibinfo {title} {Lower bound on the {Hartree-Fock} energy of the electron gas},\ }}\href {\doibase 10.1103/PhysRevA.99.052501} {\bibfield  {journal} {\bibinfo  {journal} {Phys. Rev. A}\ }\textbf {\bibinfo {volume} {99}},\ \bibinfo {pages} {052501} (\bibinfo {year} {2019})}\BibitemShut {NoStop}%
\bibitem [{\citenamefont {Christiansen}\ \emph {et~al.}(2023)\citenamefont {Christiansen}, \citenamefont {Hainzl},\ and\ \citenamefont {Nam}}]{Christiansen2023}%
  \BibitemOpen
  \bibfield  {author} {\bibinfo {author} {\bibfnamefont {M.~R.}\ \bibnamefont {Christiansen}}, \bibinfo {author} {\bibfnamefont {C.}~\bibnamefont {Hainzl}}, \ and\ \bibinfo {author} {\bibfnamefont {P.~T.}\ \bibnamefont {Nam}},\ }\bibfield  {title} {\emph {\bibinfo {title} {The {Gell-Mann–Brueckner} formula for the correlation energy of the electron gas: A rigorous upper bound in the mean-field regime},\ }}\href {\doibase 10.1007/s00220-023-04672-2} {\bibfield  {journal} {\bibinfo  {journal} {Communications in Mathematical Physics}\ }\textbf {\bibinfo {volume} {401}},\ \bibinfo {pages} {1469} (\bibinfo {year} {2023})}\BibitemShut {NoStop}%
\bibitem [{\citenamefont {Eich}\ \emph {et~al.}(2010)\citenamefont {Eich}, \citenamefont {Kurth}, \citenamefont {Proetto}, \citenamefont {Sharma},\ and\ \citenamefont {Gross}}]{FGEich_SDW_3DEG}%
  \BibitemOpen
  \bibfield  {author} {\bibinfo {author} {\bibfnamefont {F.~G.}\ \bibnamefont {Eich}}, \bibinfo {author} {\bibfnamefont {S.}~\bibnamefont {Kurth}}, \bibinfo {author} {\bibfnamefont {C.~R.}\ \bibnamefont {Proetto}}, \bibinfo {author} {\bibfnamefont {S.}~\bibnamefont {Sharma}}, \ and\ \bibinfo {author} {\bibfnamefont {E.~K.~U.}\ \bibnamefont {Gross}},\ }\bibfield  {title} {\emph {\bibinfo {title} {Noncollinear spin-spiral phase for the uniform electron gas within reduced-density-matrix-functional theory},\ }}\href {\doibase 10.1103/PhysRevB.81.024430} {\bibfield  {journal} {\bibinfo  {journal} {Phys. Rev. B}\ }\textbf {\bibinfo {volume} {81}},\ \bibinfo {pages} {024430} (\bibinfo {year} {2010})}\BibitemShut {NoStop}%
\bibitem [{\citenamefont {Baguet}\ \emph {et~al.}(2014)\citenamefont {Baguet}, \citenamefont {Delyon}, \citenamefont {Bernu},\ and\ \citenamefont {Holzmann}}]{Baguet_3DEG_2014}%
  \BibitemOpen
  \bibfield  {author} {\bibinfo {author} {\bibfnamefont {L.}~\bibnamefont {Baguet}}, \bibinfo {author} {\bibfnamefont {F.}~\bibnamefont {Delyon}}, \bibinfo {author} {\bibfnamefont {B.}~\bibnamefont {Bernu}}, \ and\ \bibinfo {author} {\bibfnamefont {M.}~\bibnamefont {Holzmann}},\ }\bibfield  {title} {\emph {\bibinfo {title} {Properties of {Hartree-Fock} solutions of the three-dimensional electron gas},\ }}\href {\doibase 10.1103/PhysRevB.90.165131} {\bibfield  {journal} {\bibinfo  {journal} {Phys. Rev. B}\ }\textbf {\bibinfo {volume} {90}},\ \bibinfo {pages} {165131} (\bibinfo {year} {2014})}\BibitemShut {NoStop}%
\bibitem [{Unl()}]{Unlike}%
  \BibitemOpen
  \href@noop {} {}\bibinfo {note} {This sailboat-like curve does not appear if choosing a perturbation of the PSP state (incoherent state) as the initial guess for the self-consistent Hartree-Fock equations, Our calculations show that the minimum total energy with respect to (pseudo)spin density wave momentum $Q$ is either approximately equal to or lower for the interlayer coherent initial guess compared to the interlayer incoherent initial guess. Hence, we present in the paper the $\varepsilon_{\rm tot}$ versus $Q/k_F$ curve using the interlayer coherent state as the initial guess. However, the sailboat-like energy curve is not a physical characteristic.}\BibitemShut {Stop}%
\bibitem [{Und()}]{Under}%
  \BibitemOpen
  \href@noop {} {}\bibinfo {note} {Under a dual gate structure, the electrostatic potential difference between the two layers also depends on the charge densities and distances to the two metallic gates. We have ignored these structural details in our calculations; therefore, our phase diagrams in Fig. 8 are only qualitatively correct.}\BibitemShut {Stop}%
\bibitem [{Com()}]{Compared}%
  \BibitemOpen
  \href@noop {} {}\bibinfo {note} {Compared to Ref. \cite{JZ_interlayerCoherence_2024}, we have an extra electrostatic energy here.}\BibitemShut {Stop}%
\bibitem [{\citenamefont {Perdew}\ and\ \citenamefont {Datta}(1980)}]{Perdew_1980}%
  \BibitemOpen
  \bibfield  {author} {\bibinfo {author} {\bibfnamefont {J.~P.}\ \bibnamefont {Perdew}}\ and\ \bibinfo {author} {\bibfnamefont {T.}~\bibnamefont {Datta}},\ }\bibfield  {title} {\emph {\bibinfo {title} {Charge and spin density waves in jellium},\ }}\href {\doibase https://doi.org/10.1002/pssb.2221020126} {\bibfield  {journal} {\bibinfo  {journal} {physica status solidi (b)}\ }\textbf {\bibinfo {volume} {102}},\ \bibinfo {pages} {283} (\bibinfo {year} {1980})}\BibitemShut {NoStop}%
\bibitem [{\citenamefont {Brener}\ and\ \citenamefont {Fry}(1981)}]{Brener_1981}%
  \BibitemOpen
  \bibfield  {author} {\bibinfo {author} {\bibfnamefont {N.~E.}\ \bibnamefont {Brener}}\ and\ \bibinfo {author} {\bibfnamefont {J.~L.}\ \bibnamefont {Fry}},\ }\bibfield  {title} {\emph {\bibinfo {title} {Spin density waves in the uniform electron gas},\ }}\href {\doibase 10.1063/1.329659} {\bibfield  {journal} {\bibinfo  {journal} {Journal of Applied Physics}\ }\textbf {\bibinfo {volume} {52}},\ \bibinfo {pages} {1624} (\bibinfo {year} {1981})}\BibitemShut {NoStop}%
\bibitem [{\citenamefont {Amusia}(1966)}]{Amusia_SDW_screen_1966}%
  \BibitemOpen
  \bibfield  {author} {\bibinfo {author} {\bibfnamefont {M.}~\bibnamefont {Amusia}},\ }\bibfield  {title} {\emph {\bibinfo {title} {On the stability of the dense electron gas},\ }}\href {\doibase https://doi.org/10.1016/0031-9163(66)91131-0} {\bibfield  {journal} {\bibinfo  {journal} {Physics Letters}\ }\textbf {\bibinfo {volume} {20}},\ \bibinfo {pages} {596} (\bibinfo {year} {1966})}\BibitemShut {NoStop}%
\bibitem [{\citenamefont {Wolff}(1960)}]{Wolff_RPA_1960}%
  \BibitemOpen
  \bibfield  {author} {\bibinfo {author} {\bibfnamefont {P.~A.}\ \bibnamefont {Wolff}},\ }\bibfield  {title} {\emph {\bibinfo {title} {Spin susceptibility of an electron gas},\ }}\href {\doibase 10.1103/PhysRev.120.814} {\bibfield  {journal} {\bibinfo  {journal} {Phys. Rev.}\ }\textbf {\bibinfo {volume} {120}},\ \bibinfo {pages} {814} (\bibinfo {year} {1960})}\BibitemShut {NoStop}%
\bibitem [{\citenamefont {Hamann}\ and\ \citenamefont {Overhauser}(1966)}]{Hamann_correlation_1966}%
  \BibitemOpen
  \bibfield  {author} {\bibinfo {author} {\bibfnamefont {D.~R.}\ \bibnamefont {Hamann}}\ and\ \bibinfo {author} {\bibfnamefont {A.~W.}\ \bibnamefont {Overhauser}},\ }\bibfield  {title} {\emph {\bibinfo {title} {Electron-gas spin susceptibility},\ }}\href {\doibase 10.1103/PhysRev.143.183} {\bibfield  {journal} {\bibinfo  {journal} {Phys. Rev.}\ }\textbf {\bibinfo {volume} {143}},\ \bibinfo {pages} {183} (\bibinfo {year} {1966})}\BibitemShut {NoStop}%
\bibitem [{\citenamefont {Sreejith}\ \emph {et~al.}()\citenamefont {Sreejith}, \citenamefont {Sau},\ and\ \citenamefont {Das~Sarma}}]{Sreejith_2024}%
  \BibitemOpen
  \bibfield  {author} {\bibinfo {author} {\bibfnamefont {G.~J.}\ \bibnamefont {Sreejith}}, \bibinfo {author} {\bibfnamefont {J.~D.}\ \bibnamefont {Sau}}, \ and\ \bibinfo {author} {\bibfnamefont {S.}~\bibnamefont {Das~Sarma}},\ }\bibfield  {title} {\emph {\bibinfo {title} {Eliashberg theory for bilayer exciton condensation},\ }}\href {https://arxiv.org/abs/2401.12313} {\ }\Eprint {http://arxiv.org/abs/arXiv:2401.12313}{arXiv:2401.12313}\BibitemShut {NoStop}%
\bibitem [{\citenamefont {Das~Sarma}\ \emph {et~al.}(1990)\citenamefont {Das~Sarma}, \citenamefont {Jain},\ and\ \citenamefont {Jalabert}}]{SDS_screen_1990}%
  \BibitemOpen
  \bibfield  {author} {\bibinfo {author} {\bibfnamefont {S.}~\bibnamefont {Das~Sarma}}, \bibinfo {author} {\bibfnamefont {J.~K.}\ \bibnamefont {Jain}}, \ and\ \bibinfo {author} {\bibfnamefont {R.}~\bibnamefont {Jalabert}},\ }\bibfield  {title} {\emph {\bibinfo {title} {Many-body theory of energy relaxation in an excited-electron gas via optical-phonon emission},\ }}\href {\doibase 10.1103/PhysRevB.41.3561} {\bibfield  {journal} {\bibinfo  {journal} {Phys. Rev. B}\ }\textbf {\bibinfo {volume} {41}},\ \bibinfo {pages} {3561} (\bibinfo {year} {1990})}\BibitemShut {NoStop}%
\bibitem [{\citenamefont {Kim}\ and\ \citenamefont {Schwartz}(1972)}]{DJKim_negative_epsilon_1972}%
  \BibitemOpen
  \bibfield  {author} {\bibinfo {author} {\bibfnamefont {D.~J.}\ \bibnamefont {Kim}}\ and\ \bibinfo {author} {\bibfnamefont {B.~B.}\ \bibnamefont {Schwartz}},\ }\bibfield  {title} {\emph {\bibinfo {title} {Negative dielectric constant of the ferromagnetic electron gas},\ }}\href {\doibase 10.1103/PhysRevLett.28.310} {\bibfield  {journal} {\bibinfo  {journal} {Phys. Rev. Lett.}\ }\textbf {\bibinfo {volume} {28}},\ \bibinfo {pages} {310} (\bibinfo {year} {1972})}\BibitemShut {NoStop}%
\bibitem [{\citenamefont {Zhang}\ \emph {et~al.}(2011)\citenamefont {Zhang}, \citenamefont {Jung}, \citenamefont {Fiete}, \citenamefont {Niu},\ and\ \citenamefont {MacDonald}}]{FZhang_pseudospin_2011}%
  \BibitemOpen
  \bibfield  {author} {\bibinfo {author} {\bibfnamefont {F.}~\bibnamefont {Zhang}}, \bibinfo {author} {\bibfnamefont {J.}~\bibnamefont {Jung}}, \bibinfo {author} {\bibfnamefont {G.~A.}\ \bibnamefont {Fiete}}, \bibinfo {author} {\bibfnamefont {Q.}~\bibnamefont {Niu}}, \ and\ \bibinfo {author} {\bibfnamefont {A.~H.}\ \bibnamefont {MacDonald}},\ }\bibfield  {title} {\emph {\bibinfo {title} {Spontaneous quantum {Hall} states in chirally stacked few-layer graphene systems},\ }}\href {\doibase 10.1103/PhysRevLett.106.156801} {\bibfield  {journal} {\bibinfo  {journal} {Phys. Rev. Lett.}\ }\textbf {\bibinfo {volume} {106}},\ \bibinfo {pages} {156801} (\bibinfo {year} {2011})}\BibitemShut {NoStop}%
\bibitem [{\citenamefont {Giuliani}\ and\ \citenamefont {Quinn}(1985{\natexlab{a}})}]{Giuliani_SDW_LL_1985}%
  \BibitemOpen
  \bibfield  {author} {\bibinfo {author} {\bibfnamefont {G.~F.}\ \bibnamefont {Giuliani}}\ and\ \bibinfo {author} {\bibfnamefont {J.~J.}\ \bibnamefont {Quinn}},\ }\bibfield  {title} {\emph {\bibinfo {title} {Spin-polarization instability in a tilted magnetic field of a two-dimensional electron gas with filled {Landau} levels},\ }}\href {\doibase 10.1103/PhysRevB.31.6228} {\bibfield  {journal} {\bibinfo  {journal} {Phys. Rev. B}\ }\textbf {\bibinfo {volume} {31}},\ \bibinfo {pages} {6228} (\bibinfo {year} {1985}{\natexlab{a}})}\BibitemShut {NoStop}%
\bibitem [{\citenamefont {Giuliani}\ and\ \citenamefont {Quinn}(1986)}]{Giuliani_SDW_LL_1986}%
  \BibitemOpen
  \bibfield  {author} {\bibinfo {author} {\bibfnamefont {G.~F.}\ \bibnamefont {Giuliani}}\ and\ \bibinfo {author} {\bibfnamefont {J.}~\bibnamefont {Quinn}},\ }\bibfield  {title} {\emph {\bibinfo {title} {Magnetic instabilities of a two-dimensional electron gas in a large magnetic field},\ }}\href {\doibase https://doi.org/10.1016/0039-6028(86)90981-7} {\bibfield  {journal} {\bibinfo  {journal} {Surface Science}\ }\textbf {\bibinfo {volume} {170}},\ \bibinfo {pages} {316} (\bibinfo {year} {1986})}\BibitemShut {NoStop}%
\bibitem [{\citenamefont {Giuliani}\ and\ \citenamefont {Quinn}(1985{\natexlab{b}})}]{Giuliani_exciton_LL_1985}%
  \BibitemOpen
  \bibfield  {author} {\bibinfo {author} {\bibfnamefont {G.}~\bibnamefont {Giuliani}}\ and\ \bibinfo {author} {\bibfnamefont {J.}~\bibnamefont {Quinn}},\ }\bibfield  {title} {\emph {\bibinfo {title} {Triplet exciton and ferromagnetic instability of a two-dimensional electron gas in a large magnetic field with filling factor $\nu$=2},\ }}\href {\doibase https://doi.org/10.1016/0038-1098(85)90176-0} {\bibfield  {journal} {\bibinfo  {journal} {Solid State Communications}\ }\textbf {\bibinfo {volume} {54}},\ \bibinfo {pages} {1013} (\bibinfo {year} {1985}{\natexlab{b}})}\BibitemShut {NoStop}%
\bibitem [{\citenamefont {Piazza}\ \emph {et~al.}(1999)\citenamefont {Piazza}, \citenamefont {Pellegrini}, \citenamefont {Beltram}, \citenamefont {Wegscheider}, \citenamefont {Jungwirth},\ and\ \citenamefont {MacDonald}}]{Piazza_SDW_LL_1999}%
  \BibitemOpen
  \bibfield  {author} {\bibinfo {author} {\bibfnamefont {V.}~\bibnamefont {Piazza}}, \bibinfo {author} {\bibfnamefont {V.}~\bibnamefont {Pellegrini}}, \bibinfo {author} {\bibfnamefont {F.}~\bibnamefont {Beltram}}, \bibinfo {author} {\bibfnamefont {W.}~\bibnamefont {Wegscheider}}, \bibinfo {author} {\bibfnamefont {T.}~\bibnamefont {Jungwirth}}, \ and\ \bibinfo {author} {\bibfnamefont {A.~H.}\ \bibnamefont {MacDonald}},\ }\bibfield  {title} {\emph {\bibinfo {title} {First-order phase transitions in a quantum {Hall} ferromagnet},\ }}\href {\doibase 10.1038/45189} {\bibfield  {journal} {\bibinfo  {journal} {Nature}\ }\textbf {\bibinfo {volume} {402}},\ \bibinfo {pages} {638} (\bibinfo {year} {1999})}\BibitemShut {NoStop}%
\bibitem [{\citenamefont {Marinescu}\ \emph {et~al.}(2000)\citenamefont {Marinescu}, \citenamefont {Quinn},\ and\ \citenamefont {Giuliani}}]{Marinescu_SDW_2000}%
  \BibitemOpen
  \bibfield  {author} {\bibinfo {author} {\bibfnamefont {D.~C.}\ \bibnamefont {Marinescu}}, \bibinfo {author} {\bibfnamefont {J.~J.}\ \bibnamefont {Quinn}}, \ and\ \bibinfo {author} {\bibfnamefont {G.~F.}\ \bibnamefont {Giuliani}},\ }\bibfield  {title} {\emph {\bibinfo {title} {Spin instabilities in semiconductor superlattices},\ }}\href {\doibase 10.1103/PhysRevB.61.7245} {\bibfield  {journal} {\bibinfo  {journal} {Phys. Rev. B}\ }\textbf {\bibinfo {volume} {61}},\ \bibinfo {pages} {7245} (\bibinfo {year} {2000})}\BibitemShut {NoStop}%
\bibitem [{\citenamefont {Yoshizawa}\ and\ \citenamefont {Takayanagi}(2007)}]{Yoshizawa_SDW_2007}%
  \BibitemOpen
  \bibfield  {author} {\bibinfo {author} {\bibfnamefont {K.}~\bibnamefont {Yoshizawa}}\ and\ \bibinfo {author} {\bibfnamefont {K.}~\bibnamefont {Takayanagi}},\ }\bibfield  {title} {\emph {\bibinfo {title} {Spin instability of integer quantum {Hall} systems},\ }}\href {\doibase 10.1103/PhysRevB.76.155329} {\bibfield  {journal} {\bibinfo  {journal} {Phys. Rev. B}\ }\textbf {\bibinfo {volume} {76}},\ \bibinfo {pages} {155329} (\bibinfo {year} {2007})}\BibitemShut {NoStop}%
\bibitem [{\citenamefont {Yoshizawa}\ and\ \citenamefont {Takayanagi}(2009)}]{Yoshizawa_SDW_QH_2009}%
  \BibitemOpen
  \bibfield  {author} {\bibinfo {author} {\bibfnamefont {K.}~\bibnamefont {Yoshizawa}}\ and\ \bibinfo {author} {\bibfnamefont {K.}~\bibnamefont {Takayanagi}},\ }\bibfield  {title} {\emph {\bibinfo {title} {Magnetic phase diagram of $\ensuremath{\nu}=2$ quantum {Hall} systems},\ }}\href {\doibase 10.1103/PhysRevB.79.125321} {\bibfield  {journal} {\bibinfo  {journal} {Phys. Rev. B}\ }\textbf {\bibinfo {volume} {79}},\ \bibinfo {pages} {125321} (\bibinfo {year} {2009})}\BibitemShut {NoStop}%
\bibitem [{\citenamefont {Zheng}\ \emph {et~al.}(2011)\citenamefont {Zheng}, \citenamefont {Marinescu},\ and\ \citenamefont {Giuliani}}]{Zheng_SDW_2011}%
  \BibitemOpen
  \bibfield  {author} {\bibinfo {author} {\bibfnamefont {L.}~\bibnamefont {Zheng}}, \bibinfo {author} {\bibfnamefont {D.~C.}\ \bibnamefont {Marinescu}}, \ and\ \bibinfo {author} {\bibfnamefont {G.~F.}\ \bibnamefont {Giuliani}},\ }\bibfield  {title} {\emph {\bibinfo {title} {Spin density waves in a semiconductor superlattice in a tilted magnetic field},\ }}\href {\doibase 10.1103/PhysRevB.84.205321} {\bibfield  {journal} {\bibinfo  {journal} {Phys. Rev. B}\ }\textbf {\bibinfo {volume} {84}},\ \bibinfo {pages} {205321} (\bibinfo {year} {2011})}\BibitemShut {NoStop}%
\bibitem [{\citenamefont {Bultinck}\ \emph {et~al.}(2020)\citenamefont {Bultinck}, \citenamefont {Khalaf}, \citenamefont {Liu}, \citenamefont {Chatterjee}, \citenamefont {Vishwanath},\ and\ \citenamefont {Zaletel}}]{Bultinck_hidden_2020}%
  \BibitemOpen
  \bibfield  {author} {\bibinfo {author} {\bibfnamefont {N.}~\bibnamefont {Bultinck}}, \bibinfo {author} {\bibfnamefont {E.}~\bibnamefont {Khalaf}}, \bibinfo {author} {\bibfnamefont {S.}~\bibnamefont {Liu}}, \bibinfo {author} {\bibfnamefont {S.}~\bibnamefont {Chatterjee}}, \bibinfo {author} {\bibfnamefont {A.}~\bibnamefont {Vishwanath}}, \ and\ \bibinfo {author} {\bibfnamefont {M.~P.}\ \bibnamefont {Zaletel}},\ }\bibfield  {title} {\emph {\bibinfo {title} {Ground state and hidden symmetry of magic-angle graphene at even integer filling},\ }}\href {\doibase 10.1103/PhysRevX.10.031034} {\bibfield  {journal} {\bibinfo  {journal} {Phys. Rev. X}\ }\textbf {\bibinfo {volume} {10}},\ \bibinfo {pages} {031034} (\bibinfo {year} {2020})}\BibitemShut {NoStop}%
\bibitem [{\citenamefont {Po}\ \emph {et~al.}(2018)\citenamefont {Po}, \citenamefont {Zou}, \citenamefont {Vishwanath},\ and\ \citenamefont {Senthil}}]{Po_IVC_2018}%
  \BibitemOpen
  \bibfield  {author} {\bibinfo {author} {\bibfnamefont {H.~C.}\ \bibnamefont {Po}}, \bibinfo {author} {\bibfnamefont {L.}~\bibnamefont {Zou}}, \bibinfo {author} {\bibfnamefont {A.}~\bibnamefont {Vishwanath}}, \ and\ \bibinfo {author} {\bibfnamefont {T.}~\bibnamefont {Senthil}},\ }\bibfield  {title} {\emph {\bibinfo {title} {Origin of {Mott} insulating behavior and superconductivity in twisted bilayer graphene},\ }}\href {\doibase 10.1103/PhysRevX.8.031089} {\bibfield  {journal} {\bibinfo  {journal} {Phys. Rev. X}\ }\textbf {\bibinfo {volume} {8}},\ \bibinfo {pages} {031089} (\bibinfo {year} {2018})}\BibitemShut {NoStop}%
\bibitem [{\citenamefont {Kwan}\ \emph {et~al.}(2021)\citenamefont {Kwan}, \citenamefont {Wagner}, \citenamefont {Soejima}, \citenamefont {Zaletel}, \citenamefont {Simon}, \citenamefont {Parameswaran},\ and\ \citenamefont {Bultinck}}]{YHKwan_kekule_2021}%
  \BibitemOpen
  \bibfield  {author} {\bibinfo {author} {\bibfnamefont {Y.~H.}\ \bibnamefont {Kwan}}, \bibinfo {author} {\bibfnamefont {G.}~\bibnamefont {Wagner}}, \bibinfo {author} {\bibfnamefont {T.}~\bibnamefont {Soejima}}, \bibinfo {author} {\bibfnamefont {M.~P.}\ \bibnamefont {Zaletel}}, \bibinfo {author} {\bibfnamefont {S.~H.}\ \bibnamefont {Simon}}, \bibinfo {author} {\bibfnamefont {S.~A.}\ \bibnamefont {Parameswaran}}, \ and\ \bibinfo {author} {\bibfnamefont {N.}~\bibnamefont {Bultinck}},\ }\bibfield  {title} {\emph {\bibinfo {title} {Kekul\'e spiral order at all nonzero integer fillings in twisted bilayer graphene},\ }}\href {\doibase 10.1103/PhysRevX.11.041063} {\bibfield  {journal} {\bibinfo  {journal} {Phys. Rev. X}\ }\textbf {\bibinfo {volume} {11}},\ \bibinfo {pages} {041063} (\bibinfo {year} {2021})}\BibitemShut {NoStop}%
\bibitem [{\citenamefont {Chatterjee}\ \emph {et~al.}(2022)\citenamefont {Chatterjee}, \citenamefont {Wang}, \citenamefont {Berg},\ and\ \citenamefont {Zaletel}}]{Chatterjee_IVC_2022}%
  \BibitemOpen
  \bibfield  {author} {\bibinfo {author} {\bibfnamefont {S.}~\bibnamefont {Chatterjee}}, \bibinfo {author} {\bibfnamefont {T.}~\bibnamefont {Wang}}, \bibinfo {author} {\bibfnamefont {E.}~\bibnamefont {Berg}}, \ and\ \bibinfo {author} {\bibfnamefont {M.~P.}\ \bibnamefont {Zaletel}},\ }\bibfield  {title} {\emph {\bibinfo {title} {Inter-valley coherent order and isospin fluctuation mediated superconductivity in rhombohedral trilayer graphene},\ }}\href {\doibase 10.1038/s41467-022-33561-w} {\bibfield  {journal} {\bibinfo  {journal} {Nature Communications}\ }\textbf {\bibinfo {volume} {13}},\ \bibinfo {pages} {6013} (\bibinfo {year} {2022})}\BibitemShut {NoStop}%
\bibitem [{\citenamefont {Xie}\ and\ \citenamefont {Das~Sarma}(2023)}]{MXie_SOC_BLG_2023}%
  \BibitemOpen
  \bibfield  {author} {\bibinfo {author} {\bibfnamefont {M.}~\bibnamefont {Xie}}\ and\ \bibinfo {author} {\bibfnamefont {S.}~\bibnamefont {Das~Sarma}},\ }\bibfield  {title} {\emph {\bibinfo {title} {Flavor symmetry breaking in spin-orbit coupled bilayer graphene},\ }}\href {\doibase 10.1103/PhysRevB.107.L201119} {\bibfield  {journal} {\bibinfo  {journal} {Phys. Rev. B}\ }\textbf {\bibinfo {volume} {107}},\ \bibinfo {pages} {L201119} (\bibinfo {year} {2023})}\BibitemShut {NoStop}%
\bibitem [{\citenamefont {Das}\ and\ \citenamefont {Huang}(2024)}]{CHuang_QuarterMetal_2023}%
  \BibitemOpen
  \bibfield  {author} {\bibinfo {author} {\bibfnamefont {M.}~\bibnamefont {Das}}\ and\ \bibinfo {author} {\bibfnamefont {C.}~\bibnamefont {Huang}},\ }\bibfield  {title} {\emph {\bibinfo {title} {Quarter-metal phases in multilayer graphene: {Ising-XY} and annular {Lifshitz} transitions},\ }}\href {\doibase 10.1103/PhysRevB.110.035103} {\bibfield  {journal} {\bibinfo  {journal} {Phys. Rev. B}\ }\textbf {\bibinfo {volume} {110}},\ \bibinfo {pages} {035103} (\bibinfo {year} {2024})}\BibitemShut {NoStop}%
\bibitem [{\citenamefont {Lu}\ \emph {et~al.}(2022)\citenamefont {Lu}, \citenamefont {Wang}, \citenamefont {Chatterjee},\ and\ \citenamefont {You}}]{DCLu_RTG_2022}%
  \BibitemOpen
  \bibfield  {author} {\bibinfo {author} {\bibfnamefont {D.-C.}\ \bibnamefont {Lu}}, \bibinfo {author} {\bibfnamefont {T.}~\bibnamefont {Wang}}, \bibinfo {author} {\bibfnamefont {S.}~\bibnamefont {Chatterjee}}, \ and\ \bibinfo {author} {\bibfnamefont {Y.-Z.}\ \bibnamefont {You}},\ }\bibfield  {title} {\emph {\bibinfo {title} {Correlated metals and unconventional superconductivity in rhombohedral trilayer graphene: A renormalization group analysis},\ }}\href {\doibase 10.1103/PhysRevB.106.155115} {\bibfield  {journal} {\bibinfo  {journal} {Phys. Rev. B}\ }\textbf {\bibinfo {volume} {106}},\ \bibinfo {pages} {155115} (\bibinfo {year} {2022})}\BibitemShut {NoStop}%
\bibitem [{\citenamefont {Huang}\ \emph {et~al.}(2023)\citenamefont {Huang}, \citenamefont {Wolf}, \citenamefont {Qin}, \citenamefont {Wei}, \citenamefont {Blinov},\ and\ \citenamefont {MacDonald}}]{CHuang_RTG_2023}%
  \BibitemOpen
  \bibfield  {author} {\bibinfo {author} {\bibfnamefont {C.}~\bibnamefont {Huang}}, \bibinfo {author} {\bibfnamefont {T.~M.~R.}\ \bibnamefont {Wolf}}, \bibinfo {author} {\bibfnamefont {W.}~\bibnamefont {Qin}}, \bibinfo {author} {\bibfnamefont {N.}~\bibnamefont {Wei}}, \bibinfo {author} {\bibfnamefont {I.~V.}\ \bibnamefont {Blinov}}, \ and\ \bibinfo {author} {\bibfnamefont {A.~H.}\ \bibnamefont {MacDonald}},\ }\bibfield  {title} {\emph {\bibinfo {title} {Spin and orbital metallic magnetism in rhombohedral trilayer graphene},\ }}\href {\doibase 10.1103/PhysRevB.107.L121405} {\bibfield  {journal} {\bibinfo  {journal} {Phys. Rev. B}\ }\textbf {\bibinfo {volume} {107}},\ \bibinfo {pages} {L121405} (\bibinfo {year} {2023})}\BibitemShut {NoStop}%
\end{thebibliography}

%

\end{document}